\documentclass[conference]{IEEEtran}

\usepackage[table,xcdraw]{xcolor}
\usepackage{graphicx}
\usepackage{amsmath}
\usepackage[]{subfig}
\usepackage{float}
\usepackage[T1]{fontenc}
\usepackage{array}
\usepackage{hyperref}
\newcolumntype{L}[1]{>{\centering\arraybackslash}m{#1}}
\usepackage{etoolbox}
\usepackage{xspace}
\usepackage{makecell}
\usepackage{lipsum}
\usepackage{booktabs}
\usepackage{tabularx}
\usepackage{graphics}
\usepackage{multirow}
\usepackage{enumitem}

\newcommand{\pl}{Peerlock}
\begin{document}

\title{Flexsealing BGP Against Route Leaks:\\ Peerlock Active Measurement and Analysis}

% author names and affiliations
% use a multiple column layout for up to three different
% affiliations
\author{\IEEEauthorblockN{Tyler McDaniel, Jared M. Smith, and Max Schuchard}
\IEEEauthorblockA{University of Tennessee, Knoxville\\
\{bmcdan16, jms, mschucha\}@utk.edu}}

\IEEEoverridecommandlockouts
\makeatletter\def\@IEEEpubidpullup{6.5\baselineskip}\makeatother
\IEEEpubid{\parbox{\columnwidth}{
    Network and Distributed Systems Security (NDSS) Symposium 2021\\
   21-24 February 2021, San Diego, CA, USA\\
   ISBN 1-891562-66-5\\
    https://dx.doi.org/10.14722/ndss.2021.23080\\
    www.ndss-symposium.org
}
\hspace{\columnsep}\makebox[\columnwidth]{}}

\maketitle

\begin{abstract}
BGP route leaks frequently precipitate serious disruptions to inter-domain routing. These incidents have plagued the Internet for decades while deployment and usability issues cripple efforts to mitigate the problem. \pl{}, presented in 2016, addresses route leaks with a new approach. \pl{} enables filtering agreements between transit providers to protect their own networks without the need for broad cooperation or a trust infrastructure. We outline the \pl{} system and one variant, Peerlock-lite, and conduct live Internet experiments to measure their deployment on the control plane. Our measurements find evidence for significant \pl{} protection between Tier 1 networks in the peering clique, where 48\% of potential \pl{} filters are deployed, and reveal that many other networks also deploy filters against Tier 1 leaks. To guide further deployment, we also quantify \pl{}'s impact on route leaks both at currently observed levels and under hypothetical future deployment scenarios via BGP simulation. These experiments reveal present \pl{} deployment restricts Tier 1 leak export to 10\% or fewer networks for 40\% of simulated leaks. Strategic additional Peerlock-lite deployment at all large ISPs (<1\% of all networks), in tandem with \pl{} within the peering clique as deployed, completely mitigates about 80\% of simulated Tier 1 route leaks.

\end{abstract}

\section{Introduction}{\label{intro}}The Internet consists of many \textit{Autonomous Systems} (ASes) with distinct IP prefixes, routing policies, and inter-AS connections. These networks exchange routes with neighboring ASes over the \textit{control plane} to connect hosts in disparate ASes and create the illusion for users of a single, unified Internet. Unfortunately, there are few security controls on route exchange. ASes behaving adversarially, whether intentionally or by mistake, can export routes that should be kept internally or shared with only a subset of their neighbors. Because the language ASes use to communicate - the \textit{Border Gateway Protocol} or BGP - does not package validation information with routes, remote networks often receive and propagate these \textit{route leaks} throughout the control plane. Leaks frequently steer user traffic on the \textit{data plane} onto unintended paths that lack capacity for the additional traffic. The end result is soaring latency or complete availability loss for destination services. Recent route leaks to prefixes hosting major content/service providers like Spotify~\cite{enzu-leak}, Cloudflare~\cite{verizon-cloudflareleak} and Google~\cite{google-leak} have highlighted the global impact of this problem.

Existing tools designed to curtail leaks, like the many Internet Routing Registries (IRRs), are challenging to deploy or limited in scope. IRRs are databases where ASes can publish their routing policies. Other ASes can then convert IRR-stored policies into filters to validate received routes. IRR-based filtering is limited by its requirement for broad AS participation, however, as the motivations and sophistication of network operators varies greatly between ASes~\cite{kuerbis2017internet}. Other BGP security extensions, like the Resource Public Key Infrastructure (RPKI), only enable filtering for a subset of leaks (e.g. re-origination leaks for RPKI).

The \textit{\pl{}}~\cite{peerlock-talk, peerlock} leak defense system was presented in 2016 to address the need for a deployable solution. Each \pl{} deployment occurs between two neighboring ASes, the \textit{protector} AS and \textit{protected} AS. The protector AS agrees to filter routes that transit the protected AS unless they arrive directly from the protected AS or one of its designated upstreams. The filter prevents the protector AS from propagating or steering its traffic onto any leaked route that transits the protected AS, regardless of origin AS/destination prefix. \pl{} is designed to leverage the rich web of relationships that exist between transit networks in the Internet's core, and functions without coordination with other ASes on potential leak paths. This makes \pl{} especially viable in the \textit{peering clique} formed by the 19 \textit{Tier 1} ASes that sit atop the inter-domain routing hierarchy. A related technique, Peerlock-lite, enables networks to spot likely leaks without prior out-of-band communication. ASes deploying Peerlock-lite drop routes arriving from customers that contain a Tier 1 AS; it is highly improbable that customers are providing transit for large global networks.

Our first contribution is a measurement of \pl{}/Peerlock-lite deployment on the control plane. In Section~\ref{measuring} we design, execute, and evaluate active Internet measurements to search for evidence of filtering consistent with these systems. Our experiments use BGP poisoning, a technique used in prior work for traffic engineering~\cite{smith2018routing} and path discovery~\cite{anwar2015investigating}, to mimic route leaks that transit some target AS. We then listen for which networks propagate - or filter - these "leaks" relative to control advertisements. This information feeds several inference techniques designed to uncover which ASes are \pl{}ing for (protecting) the target AS.

Notably, we find substantial \pl{} deployment within the peering clique: about 48\% of possible filtering rules (153/342) are already implemented within this set. Further, many non-Tier 1 ASes - including nearly 40\% of large ISPs observed during our experiments - perform some Peerlock-lite filtering on Tier 1 AS leaks. Evidence for \pl{} filtering of non-Tier 1 leaks is virtually nonexistent, though three Tier 1 networks (AS 12956, AS 2914, and AS 3320) each filter leaks for more than 20 non-Tier 1 ASes.

After detecting current \pl{}/Peerlock-lite deployments, we ask how well these systems mitigate Tier 1 leaks. Internet-scale BGP simulations in Section~\ref{simulations} test over 6,000 simulated Tier 1 leaks against observed \pl{}/Peerlock-lite levels to quantify the effect of these systems as deployed. We test the same leaks against six hypothetical extended deployment scenarios to understand where additional filters should be placed to isolate leaks. 

We find that \pl{} filtering within the peering clique is helpful, but not sufficient to mitigate Tier 1 route leaks on its own. Consistent with prior work on BGP filtering~\cite{gilad2017we}, our experiments show that positioning filters at relatively few ASes - the roughly 600 large ISPs - can play a decisive role in leak prevention. About 80\% of simulated leaks were completely mitigated by uniform Peerlock-lite filter deployment at large ISPs, with fewer than 10\% of leaks spreading beyond 10\% of the topology. These figures are especially encouraging because Peerlock-lite is based on a simple route validity check informed by the valley-free routing model~\cite{gao2001inferring} that requires no out-of-band communication.

\noindent In this paper, we make the following key contributions:
\begin{itemize}[noitemsep,topsep=0pt,parsep=0pt,partopsep=0pt]
    \item We give an overview of the \pl{} and Peerlock-lite filtering systems, and consider their benefits and limitations relative to existing tools in Section~\ref{overview}.
   \item We describe how we adapt existing Internet measurement techniques to probe \pl{}/Peerlock-lite deployment on the control plane and introduce a novel inference method in Section~\ref{method}.
  \item We actively measure where \pl{} and Peerlock-lite filters are deployed with PEERING~\cite{schlinker2019peering} and CAIDA's BGPStream~\cite{orsini2016bgpstream} in Section~\ref{inference}, with a discussion of results in Section~\ref{inference-discussion}.
   \item We simulate thousands of Tier 1 route leaks against several protection scenarios in Section~\ref{sim-eval}, and present a new path encoding method to understand how these scenarios influence leak propagation and export in Section~\ref{sim-discussion}.
\end{itemize}

\section{Background}{\label{background}}\subsection{The Border Gateway Protocol}{\label{bgp}}

The Internet is a confederation of about 69,000 smaller networks, called \textit{Autonomous Systems} or \textit{ASes}. ASes exchange routing information via the Border Gateway Protocol (BGP) to enable global connectivity. Each AS originates routes to its hosted prefixes; these routes are advertised to neighbors via \textit{BGP updates}. Each update contains a prefix and a collection of other attributes, including an AS PATH that describes the route's AS-level hops. ASes compare all received updates via the BGP decision process to select a single best path to every destination prefix. Both path qualities (like AS PATH length) and local network policies (e.g., business relationship with advertising AS) are taken in account when selecting a best path, but policies take precedence in the process. Once an AS selects a best path for a given prefix, it prepends its unique AS number (ASN) to the path and advertises \textit{only} that path to its neighbors. 

Paths learned from customer ASes - those purchasing transit - are advertised to all connections. Provider-learned routes, meanwhile, are generally only advertised to an AS's customers. Peer ASes exchange traffic without compensation, and likewise advertise routes learned from one another only to customer ASes. Limitations on non-customer learned route export prevents customer ASes from transiting traffic between peers/providers at their own expense. This dynamic, known as the Gao-Rexford or \textit{valley-free routing} model~\cite{gao2001inferring}, guides the exchange of routes on the control plane. No widely-deployed mechanism enforces this model, but the economic incentives it describes shape AS path export behavior.

The \textit{customer cone}~\cite{giotsas2014inferring} is one product of this model. An AS's customer cone is the set of all ASes reachable from the AS via only provider to customer links. Stated simply, these are the AS's direct and indirect customers. Customer cone size is one of the few publicly observable features commonly used to judge an AS's influence on the control plane, e.g. in CAIDA's AS ranking~\cite{CAIDA-rank}. Customer cone size is the basis for the UCLA classification presented in~\cite{oliveira2009completeness} widely used in research on this topic~\cite{smith2020withdrawing, tran2019feasibility, anwar2015investigating, choi2011understanding, yan2009bgpmon}. This scheme separates ASes into 1) Tier 1 ASes, who have no providers, form a peering clique, and can transit traffic to any prefix without compensation, 2) large ISPs with more than 50 customer cone ASes, 3) small ISPs with 5-50 customer cone ASes, and 4) stub ASes with fewer than 5 direct or indirect customer networks.

\subsection{Route Leaks}{\label{leak-bg}}
Despite its vital role in binding together Internet networks, BGP is missing key security features like cryptographic hardening of routes exchanged between ASes or trusted certification binding ASes to owned prefixes. This leads to two common classes of major inter-domain routing mishaps, \textit{prefix hijacking} and \textit{route leaks}. Prefix hijacks occur when a network, often unintentionally, originates or advertises a fake but attractive (e.g. shorter or more specific) route to prefixes owned by another AS. Traffic destined for those prefixes is then intercepted by the hijacker. A number of recent studies focus on hijack mitigation~\cite{sermpezis2018artemis, schlamp2016heap, wahlisch2012towards}.

Route leaks are defined in RFC 7908 as the propagation of an advertisement beyond its intended scope~\cite{leak-rfc}. Type 1-4 leaks all cover various valley-free routing violations, i.e. advertising one peer/provider's routes to another peer/provider. Because remote ASes have little or no information on relationships between non-neighboring networks, they generally cannot distinguish leaks from valid routes, and may propagate them throughout the topology. Type 5 leaks occur when one provider's routes are announced to another with the AS PATH stripped, effectively re-originating the prefix from the leaker. Finally, a Type 6 leak involves an AS announcing routes used internally to its neighbors. These routes are often more specific than externally announced routes; this makes the leaks more attractive in the BGP decision process and encourages their spread to other remote networks.

\begin{figure}[h]
	\centering
	\includegraphics[width=0.85\columnwidth]{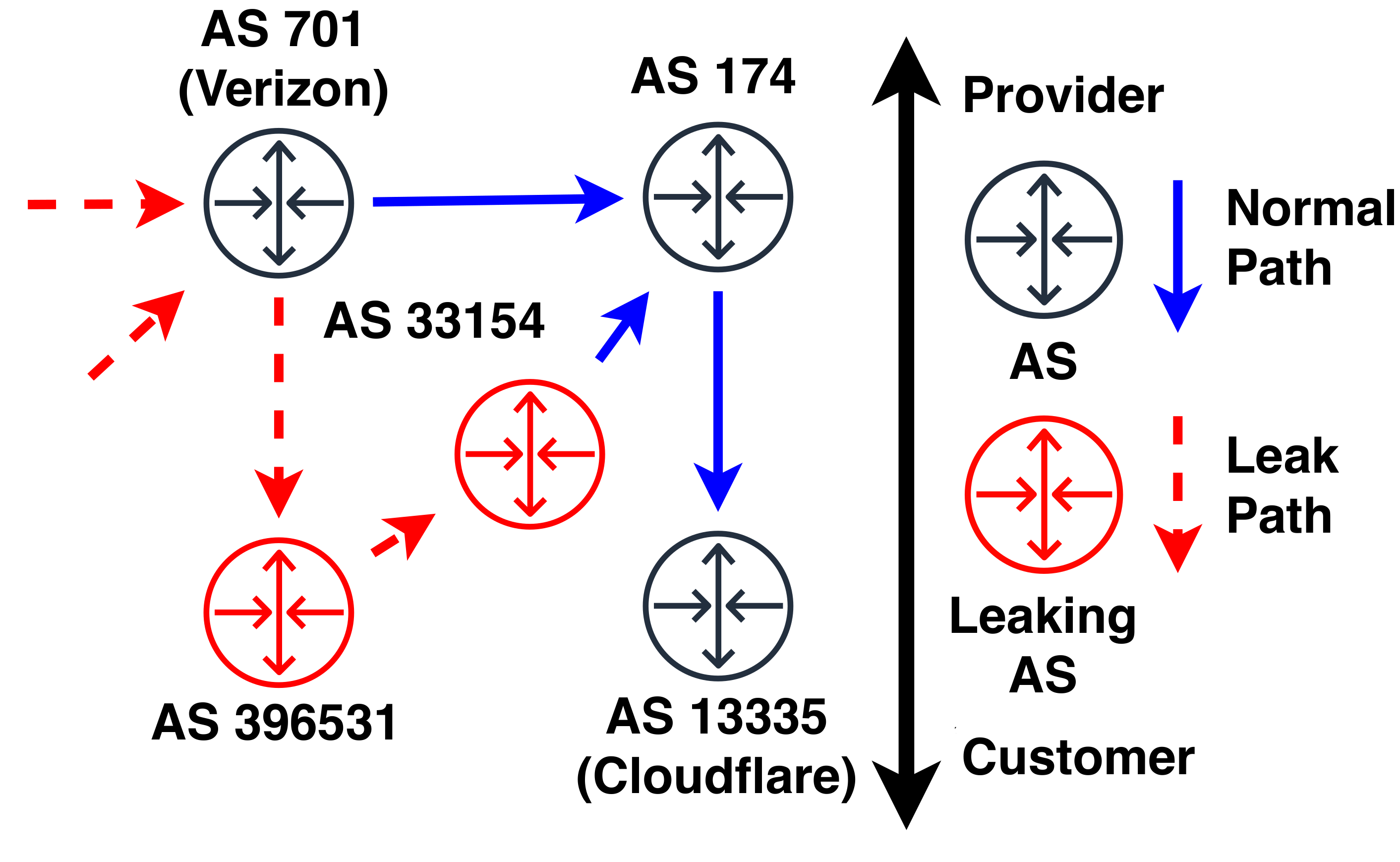}e
	\caption{2019 Verizon/Cloudflare leak. Other destination services were also affected.}
	{\label{fig:verizon_diag}}
\end{figure}

Globally disruptive route leaks occur frequently~\cite{enzu-leak, rostelecom-leak, google-leak, dyn-level3leak, indosat-leak, volumedrive-leak}. The 2019 Verizon-Cloudflare leak~\cite{verizon-cloudflareleak} is a high profile example. A small ISP, AS 33154, leaked specific internal prefixes (Type 6) to Cloudflare and many other destinations to its customer, AS 396531. AS 396531 committed a Type 1 leak by advertising this route to its other provider, AS 701 Verizon. Verizon propagated the leak, which spread widely on the control plane because it was more specific than legitimate available routes (see depiction in Fig.~\ref{fig:verizon_diag}). Traffic for Cloudflare, a leading content distribution network (CDN), was funneled through small networks. Many of the thousands of websites and services backed by Cloudflare suffered degraded service until the leak was resolved via out-of-band communication between Cloudflare and AS 33154~\cite{verizon-cloudflareleak2}.

\subsection{Route Leak Prevention}{\label{prevention}}
There are a number of tools available to assist network operators in preventing route leaks. The Resource Public Key Infrastructure~\cite{Lepinski:2012:RPKI:RFC6480} is a trusted repository for certificates that bind IP prefixes to owning ASes's public keys, called Route Origin Authorizations (ROAs). Remote networks can validate BGP updates against ROAs in the RPKI, a process called Route Origin Validation (ROV). Widespread ROV filtering could prevent Type 5 (and some Type 6) leaks and many prefix hijacking attacks. Unfortunately, ROA/ROV deployment has suffered from circular deployment logic; it is meaningless for origin ASes to invest in publishing ROAs until ROV is widely implemented, but ROV is ineffective without ROAs. This issue has been identified as a major obstacle to ROV deployment~\cite{gilad2017we, hlavacek2019disco}. NIST estimates that just 20\% of prefixes are covered by a valid ROA~\cite{rpkimon}.

Internet routing registries (IRRs) back another leak prevention system. IRRs are databases where AS operators can store their routing policies. Remote networks can ingest these policies to inform filters that block unintended/invalid advertisements. IRR databases are operated by private firms, regional Internet registries, and other interests~\cite{irr-list}, and policy entries are often mirrored between them. A complete, up-to-date IRR would eliminate many Type 1-4 route leaks. Like ROV filtering, though, IRR filtering is hampered by deployment headaches. ASes' routing policies are interdependent, so changes to one network's stored policies can render many others obsolete. Operators in smaller, resource-limited networks can avoid periodic updates by configuring permissive routing policies; large transit ASes have complex, dynamic routing policies that require frequent changes to dependent networks' filters~\cite{kuerbis2017internet}. These issues, combined with poor or non-existent authentication, have resulted in inconsistent and out-of-date IRRs. Though leading organizations like RIPE have launched efforts to improve IRR quality~\cite{ripe-cleanup}, operator incentive and dependency issues will continue to limit their usefulness.

Other filtering techniques include max-prefix limit filtering, where a network caps the number of prefixes it will accept from a neighbor. This prevents mass prefix spills like the 2017 Level 3 leak~\cite{dyn-level3leak}, but not more targeted (yet highly disruptive) leaks like the Verizon/Cloudflare incident described earlier. BGPSec~\cite{lepinski2013bgpsec} is a protocol extension for cryptographic AS path hardening. This would prevent some types of hijacking, but BGPSec has not been commercially implemented and is not designed to prevent route leaks.

Finally, a communities-based "down-only" (DO) leak solution has been proposed~\cite{down-communities}. Large BGP communities~\cite{large-communities} are signals containing three integers that can be attached to routes. The DO system relies on providers/peers marking a route "down-only" using the first two integers in a large community, with their ASN included as the third integer, before passing the route to customers or peers. If these customers/peers attempt to re-advertise the route to other providers/peers, the attached DO community will clearly signal a route leak. While this system would prevent many leaks if properly implemented, it relies on customers/peers preserving DO information when propagating advertisements. Moreover, some leaks - like the internal route leaks in the Verizon/Cloudflare incident - would not be arrested by this system.

\subsection{BGP Poisoning}
\begin{figure}[h]
	\centering
	\includegraphics[width=0.85\columnwidth]{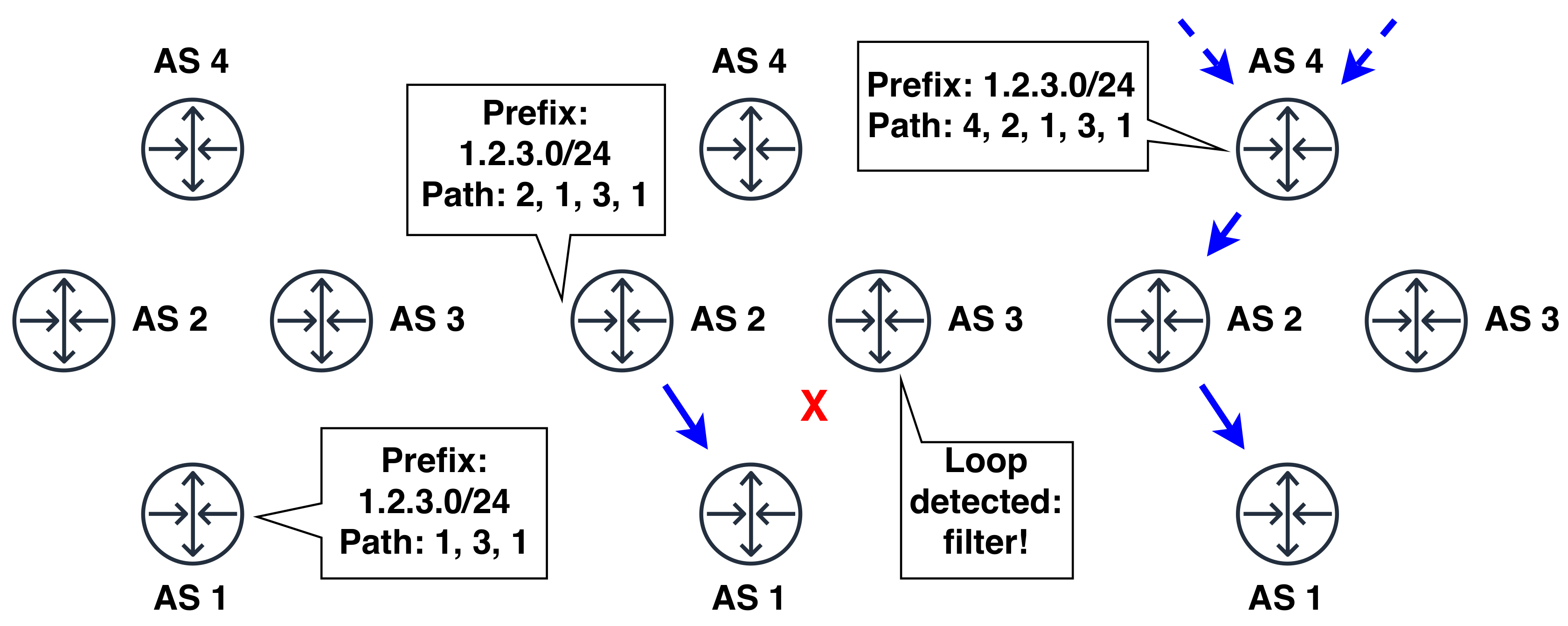}
	\caption{BGP poisoning. AS 1 originates a route with AS 2 prepended to path (left); AS 2 filters the update (center), but AS 3 propagates (right).}
	{\label{fig:poison_diag}}
\end{figure}

{\label{poisoning}}
\textit{BGP poisoning} is a technique designed to manipulate the BGP decision process in remote networks. ASes originating a prefix can \textit{poison} an advertisement by including the ASNs of remote networks in the AS PATH. Often, the poisoned ASNs will be inserted between copies of the origin's ASN. This "sandwiching" ensures traffic is routed properly and that the advertisement is valid for ROV filtering purposes (see Fig.~\ref{fig:poison_diag}). BGP prevents cycles from forming in the topology by requiring ASes to drop routes containing their own ASN in the AS PATH; this is known as BGP loop detection. So, all networks included in the poisoned update's AS PATH - the \textit{poisoned} ASes - will filter it. Poisons can be used for inbound traffic engineering purposes~\cite{smith2018routing,smith2020withdrawing,katz2012lifeguard} but we employ them in Section~\ref{measuring} to mimic route leaks transiting the poisoned AS.

\section{The \pl{} System}{\label{overview}}
The challenge of leak filtering stems from the topological scope asymmetry between BGP routes and the perspective of individual AS operators who evaluate them. Routes span the topology (global scope); operators only know their own relationships with adjoining ASes (local scope). Filtering systems built on the RPKI~\cite{huston2012validation} and IRRs~\cite{irr} partially address this asymmetry by applying additional information to the route evaluation process. However, these existing solutions have limitations that have hamstrung their deployment. Most critically, their effectiveness depends on the cooperation of many unincentivized remote ASes as detailed in Section~\ref{prevention}. 

\subsection{Peerlock}
\begin{figure}[h]
	\centering
	\includegraphics[width=0.75\columnwidth]{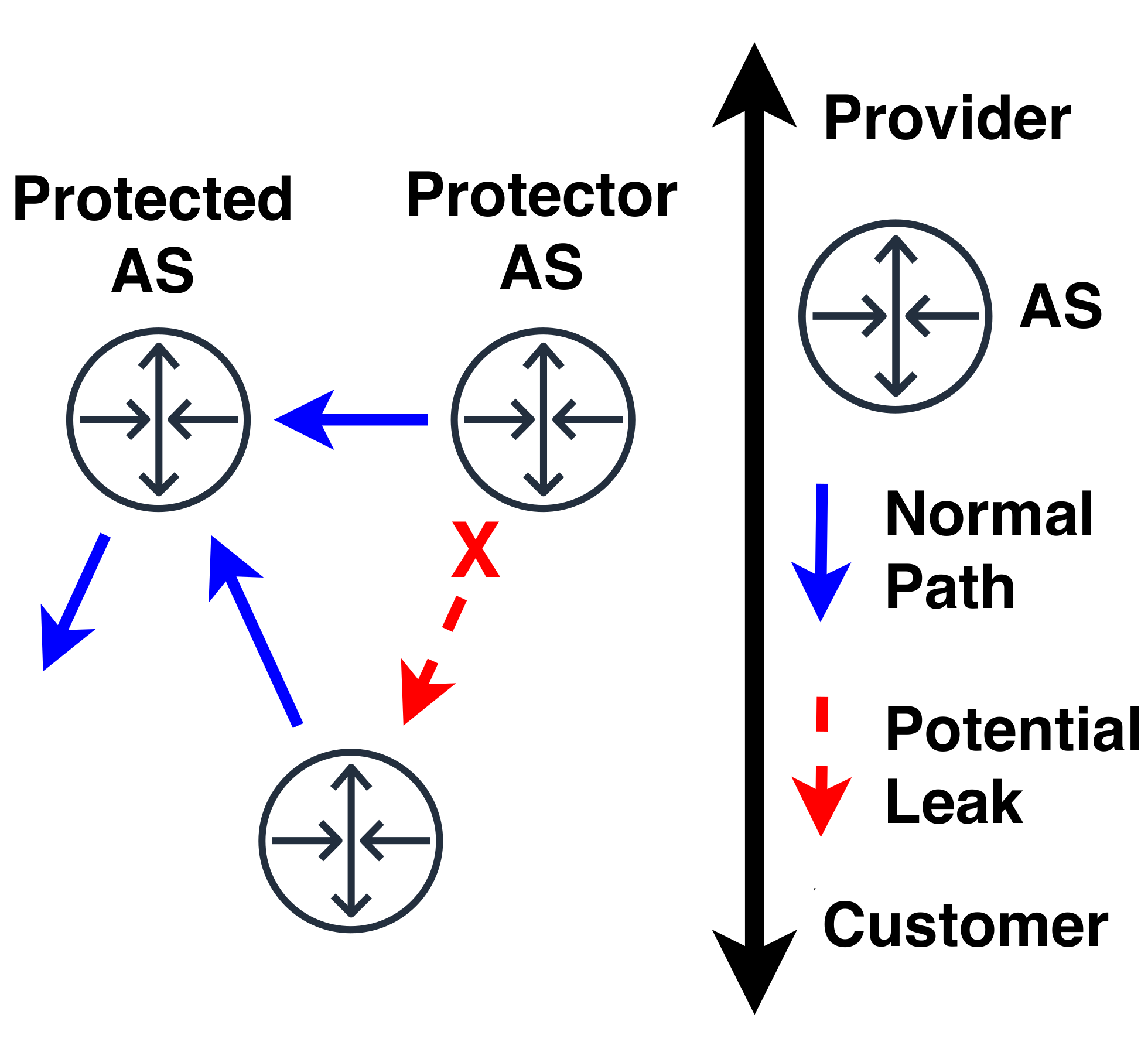}
	\caption{Simple~\pl{} deployment. Protector AS filters updates containing the peer Protected AS from unauthorized propagators.}
	{\label{fig:lock_diag}}
\end{figure}

\textit{\pl{}}, first detailed by NTT in 2016~\cite{peerlock-talk, peerlock}, is a leak filtering scheme based on out-of-band information exchange between BGP neighbors. \pl{} requires a single AS (the \textit{protected AS}) to designate authorized upstreams to their BGP neighbor (the \textit{protector AS}). This communication distributes AS relationship information between peers to decrease route/filterer scope asymmetry. The protector AS then rejects any BGP update whose AS PATH contains the protected AS unless received 1) directly from the protected AS, or 2) from an authorized upstream, with the protected AS immediately following the authorized upstream in the AS PATH. We say that the protector AS is \pl{}ing for the protected AS. See Fig.~\ref{fig:lock_diag} for a depiction of the system. In this paper, we will often refer to a single instance of \pl{} - that is, one protector/protected pairing - as a \pl{} \textit{rule}.

\begin{table*}[ht]
\begin{center}
\begin{tabular}{@{}llll@{}}
\toprule
\textbf{System} & \textbf{Coverage}                                                                                                    & \textbf{Requirements}                                                                                                   & \textbf{Notes}                                     \\ \midrule
\rowcolor[HTML]{C0C0C0} 
RPKI/ROV        & \begin{tabular}[c]{@{}l@{}}Type 5 internal leak, \\ Type 6 re-origination\end{tabular}                               & \begin{tabular}[c]{@{}l@{}}RPKI trust infrastructure, \\ local ROA registration \&  \\ remote ROV checks\end{tabular}   & Type 5 coverage depends on optional ROA max length \\
IRR Filtering   & \begin{tabular}[c]{@{}l@{}}Potentially all leak types;\\ depends on stored policy\\ object specificity.\end{tabular} & \begin{tabular}[c]{@{}l@{}}Correct, fresh policy objects \&\\ derived filters along potential leak \\ path\end{tabular} & Quality issues, misaligned incentives              \\
\rowcolor[HTML]{C0C0C0} 
Max Prefix      & All leak types                                                                                                       & \begin{tabular}[c]{@{}l@{}}Filter with meaningful max prefix \\ limit somewhere on potential leak \\ path\end{tabular}  & Only effective when many prefixes are leaked       \\ \bottomrule
\end{tabular}
\end{center}
\normalsize{Table 1: Common leak filtering systems.}
\end{table*}

Here we describe \pl{}'s benefits and drawbacks relative to previous leak prevention systems, each of which is described in detail in Section~\ref{prevention}. These comparisons are summarized in Table 1. \\

\noindent \textbf{RPKI/ROV Comparison}: \pl{} provides broader leak type coverage than RPKI/ROV filtering without a trust infrastructure requirement. However, \pl{} only applies to leaks that violate configured topological rules (Types 1-4), so Type 5 (re-origination) and Type 6 (internal route) leaks fall outside its scope. ROAs tie prefixes to valid originating ASes, so ROV filtering can prevent Type 5 leaks. Additionally, ROAs can be configured with a max prefix length to prevent some internal route leaks and hijacks, although recent work has identified vulnerabilities in this feature~\cite{gilad2017maxlength}. Because \pl{} and RPKI/ROV filtering cover different leak types, \pl{} is \textit{complementary} to ROV filtering rather than a replacement. \\

\noindent \textbf{IRR Comparison}: IRRs are policy object databases capable of storing participating networks' routing intentions with great detail and fine granularity (prefix level). Any AS wishing to enforce these intentions can automatically derive filters from stored objects using software tools, whereas \pl{} rule configuration requires setup between each protector/protected AS pair. Unfortunately, IRRs suffer from incentive misalignment, governance, and rule dependency issues as described in Section~\ref{prevention}. \pl{} rules are self-contained, and changes do not affect other rules. This encapsulation avoids the cascading dependency problem exhibited by IRRs, where one AS's policy changes may render many other AS's entries obsolete. 

Most importantly, \pl{} allows the protector AS to stop leaks that transit the protected network regardless of the actions of ASes along potential leak paths; the value of IRR-based filters depends on many remote networks to store accurate policy entries. \pl{}'s relatively light cooperation requirement only requires that ASes with an existing relationship communicate information between themselves. This dynamic enables the best resourced, positioned, and incentivized networks (i.e., those serving the most customers) to block route leak propagation regardless of other remote networks' actions. \\

\noindent \textbf{Max Prefix Comparison}: BGP's max prefix feature enables networks to limit the number of prefixes they will accept over each neighboring AS connection. Mass route leaks - those involving many prefixes - are filtered once prefix volume over an inter-AS link exceeds the preset limit. Max prefix filtering affords broad protection across leak types, but cannot stop leaks involving few (potentially critical/highly trafficked) prefixes. On the other hand, \pl{} cannot stop leaks that do not violate established topological constraints regardless of volume, but is effective against more selective leaks unprotected by max prefix limits. \\

\noindent \textbf{Other Considerations}: Currently, each \pl{} rule must be manually configured, although at least one method has been proposed to automate \pl{}~\cite{automatepl-draft}. \pl{} also lacks a standard to describe how out-of-band information is exchanged between participants. Without a detailed and secure protocol for rule configuration, \pl{} is vulnerable to exploitation; fraudulent rules affect route export, and could be used to engineer traffic flows. Furthermore, operators must define their own ad-hoc protocols for communicating rules that may not guarantee authenticity and/or confidentiality. Virtually all leak solutions discussed here, including IRR, RPKI/ROV, and AS PATH filtering, are recommended by the best practices group Mutually Agreed Norms for Routing Security (MANRS)~\cite{manrs-filter}. \\

\subsection{Peerlock-lite}
\textit{Peerlock-lite}~\cite{peerlock} (or Tier 1 filter, "big networks" filter) is a related technique, based on the assumption that transit providers should never receive a route whose AS PATH includes a Tier 1 AS from a customer. This is a valid assumption under the valley-free routing model~\cite{gao2001inferring}, because such an update implies the customer is providing transit for the Tier 1 AS; otherwise, the customer would not export (leak) the route to another provider. However, Tier 1 ASes have no providers by definition. This logic can be extended heuristically to any other large non-Tier 1 networks that the provider does not expect the customer to export.

\begin{figure}[h]
	\centering
	\includegraphics[width=0.75\columnwidth]{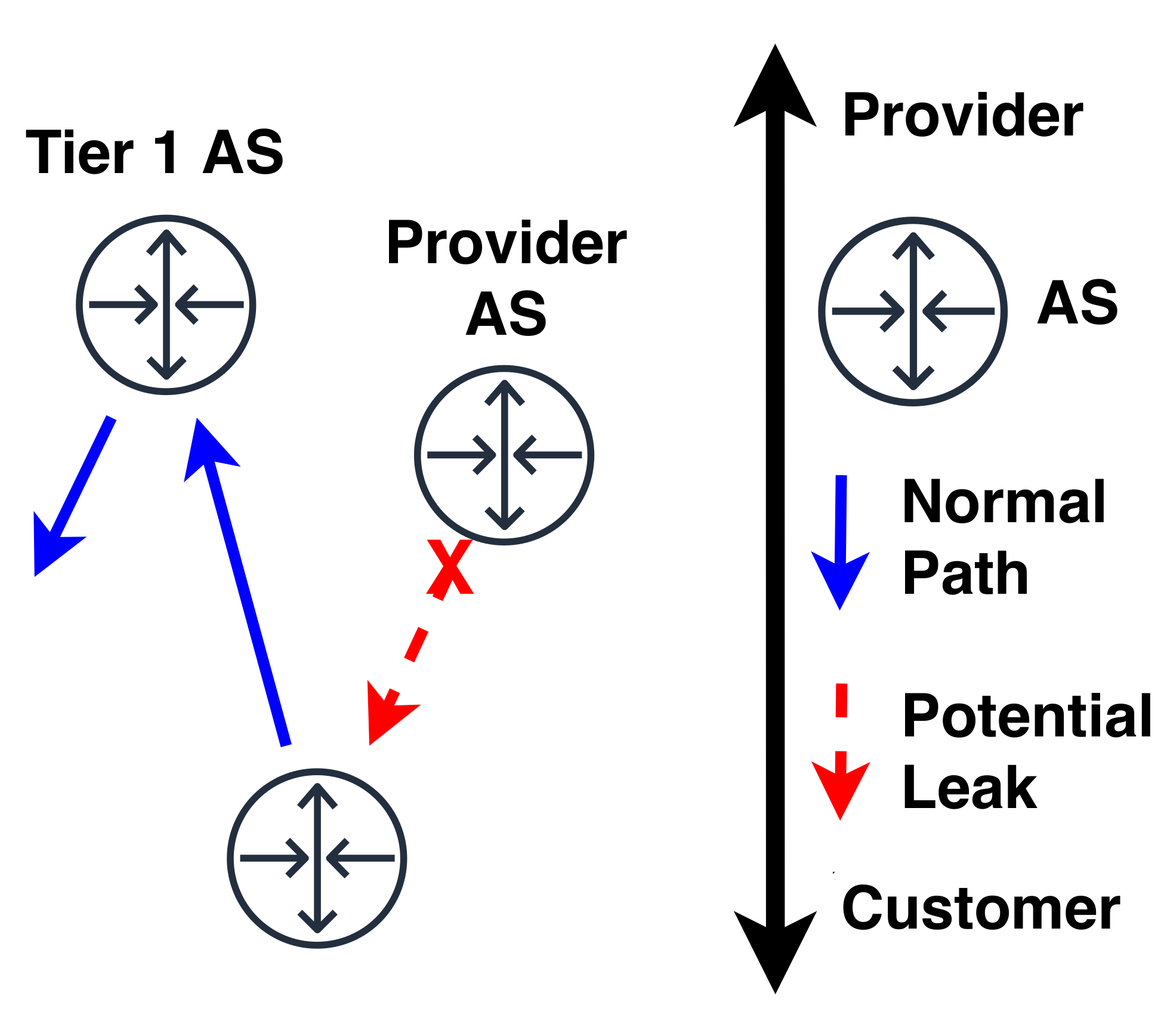}
	\caption{Example Peerlock-lite deployment. Provider AS filters updates from its customer that include a Tier 1 AS.}
	{\label{fig:lite_diag}}
\end{figure}

This simple logic yields an equally simple filtering rule for transit providers - reject any updates from customers that contain a Tier 1/large transit ASN. See Fig.~\ref{fig:lite_diag} for a depiction of this filtering technique. Peerlock-lite filters are limited to Tier 1/large transit provider leaks, but they require no out-of-band information to configure. Moreover, Tier 1 ASes' position at the Internet's core results in their frequent presence on AS PATHs of highly disruptive leaks, e.g. the Verizon/Cloudflare leak~\cite{verizon-cloudflareleak} and the Enzu/AWS/Spotify leak~\cite{enzu-leak}.

\section{Measuring \pl{} Deployment}{\label{measuring}}Our initial experiments seek to establish the current state of \pl{} deployment on the control plane. As discussed in the previous section, every \pl{} rule is configured between a pair of networks: the protector AS and the protected AS. Each of the experiments in this section works to identify some or all \pl{}/Peerlock-lite protectors for a targeted AS. 

\subsection{Measurement Methodology}{\label{method}}

\noindent\textbf{Experimental Design:}
Each set of measurement experiments in this section is designed to discover \pl{} rules for a set of potential protected ASes, called \textit{target ASes}. For each target AS, we advertise a /24 prefix from many points-of-presence (\textit{PoPs}) on the control plane. This is the \textit{control advertisement}. It is a normal /24 origination in every way, except that our university AS - which we know not to be protected by any \pl{} rule - is poisoned  (i.e., prepended to the advertisement's AS PATH - see Section~\ref{poisoning}). We then listen at varied collection sites, called \textit{collectors}, for BGP updates triggered by our advertisement. The AS PATH for each such update that arrives at collectors lists in encounter order the ASes that received and re-issued the update as described in Section~\ref{bgp}.

Taken together, the gathered AS PATHs form a directed acyclic graph (DAG) that describes the control advertisement's propagation through the control plane; each AS appearing on at least one AS PATH forms a node in the DAG, and AS ordering within paths allows us to form directed edges between nodes. BGP loop detection prevents cycles as noted in Section~\ref{poisoning}. We call this graph the \textit{control DAG}. Note that all of the ASes appearing in the control DAG propagated (and thus did not filter) control updates that include a poisoned AS. 

We then wait 30 minutes for update propagation before issuing an explicit withdrawal for the /24 prefix. This timing is built conservatively from empirical measurements of propagation times through the control plane (see update propagation experiments in appendix). After another waiting period to ensure the withdrawal has completely propagated, we issue a \textit{leak advertisement} for the same /24 prefix. This advertisement matches the control advertisement in every way, except that the target AS is poisoned. This leak advertisement structure is designed to mimic a leak for the purposes of \pl{} while avoiding other common filtering systems. The target AS's presence on update paths triggers filtering for any \pl{} protector ASes. 

Finally, we gather all BGP updates for the leak advertisement from our collectors. The ASes that appear on AS PATHs in any of these updates are added to a set called the \textit{leak set}. Since they propagated poisoned updates, we know these ASes did not filter the "leak". With the control DAG and leak set together, we can reason about which ASes are \pl{}ing for the target AS using two techniques: 1) clique inference and 2) DAG inference. 

\begin{figure}[h]
	\centering
	\includegraphics[width=0.75\columnwidth]{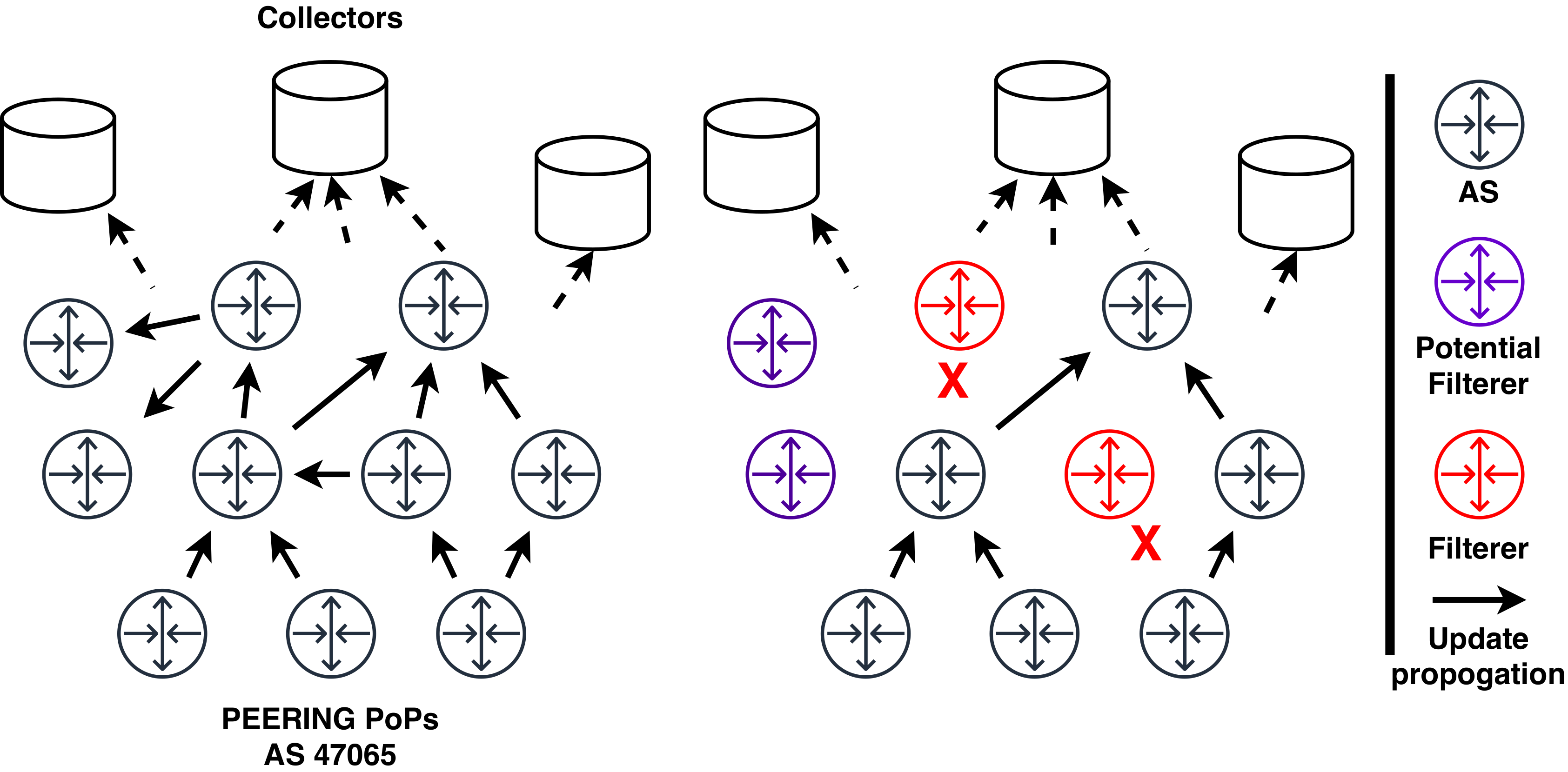}
	\caption{Measurement experiment depiction. Inferences are made about \pl{} deployment based on differences between normal updates (left) vs. poisoned updates (right) arriving at collectors.}
	{\label{fig:measure_diag}}
\end{figure}

For detecting Tier 1 protector ASes, we use \textit{clique inference}. This simple technique relies on the fact that Tier 1 ASes form a peering clique by definition. According to the valley-free routing model~\cite{gao2001inferring}, ASes share all updates received from customers with their peers; this maximizes the traffic the AS transits for its customers (and thus the AS operator's compensation). Further, ASes should not share a peer's updates with another peer, as this is a Type 2 route leak~\cite{leak-rfc}. So, in general, if a Tier 1 AS is observed propagating an update, \textit{all} Tier 1s should receive the update via their peering relationships. Because we observe at least one Tier 1 propagating control and leak updates across all experiments, we define a simple rule for inferring Tier 1 protector ASes: any Tier 1 AS that appears in the control DAG but not the leak set is \pl{}ing for the target AS.

Inferring other protector ASes requires a more general technique. Outside the structural guarantees provided by the Tier 1 clique, there is significantly more uncertainty about which networks are filtering leak updates. Specifically, it is difficult to distinguish an AS filtering updates from an AS not receiving updates at all due to filtering by other upstream/downstream networks. This challenge leads us to make three separate inferences for these ASes for each leak target. 

First and simplest is the \textit{max inference} set, defined as all control DAG ASes minus the leak set. This set includes all ASes who \textit{may have} filtered leak updates, but also ASes who did not receive the leak update because it was filtered by an intermediate AS. Secondly, we build a \textit{min inference} set. This set is built by deleting all leak set ASes from the control DAG, and collecting the root of every weakly connected component that remains. This isolates the ASes that filtered leak updates from ASes in their "shadow" who did not receive the updates. The min inference set contains those ASes who likely filtered leak updates based on routes we observed. Note that the min/max inference techniques closely align with those employed in the long path filtering experiments in Smith et al.'s study on BGP poisoning as a re-routing primitive~\cite{smith2020withdrawing}.

Our last inference set is the \textit{likely inference}. Because ASes only export their best path to our /24 prefix, we cannot observe every edge that should exist in the control DAG (i.e., every potential propagation path for updates). So, this set's is built like the min inference set, except that we augment the control DAG with edges from CAIDA's provider-peer observed customer cone inference~\cite{CAIDA}. That is, we add edges to the control DAG where CAIDA's data indicates there are links between ASes that we did not observe due to policy decisions. This forms a superset of the min inference set and a subset of the max inference set that contains the most likely filterers. This is a novel technique not used to our knowledge in any prior work on this topic.

These three inference sets are formed for each target from differences in control and leak update propagation. In addition to these sets, we also build a \textit{min/max/likely poison filtering set} by following the same steps listed above, but with a \textit{unpoisoned advertisement}'s updates compared against the control advertisement's updates. These sets are built to explore the prevalence of general poison filtering as in Smith et al.~\cite{smith2020withdrawing}.\\

\noindent\textbf{Framework Details:} The control-plane measurement framework for these experiments consists of 1) 13 PoPs to issue BGP advertisements and 2) 54 BGP collectors to listen for propagation. We employ the PEERING testbed~\cite{schlinker2019peering} for the first requirement. PEERING allows us to advertise three assigned /24 prefixes from edge routers at thirteen PoPs worldwide. For collecting BGP updates, we used CAIDA's BGPStream~\cite{CAIDA} tool. This tool draws updates from 54 globally distributed collectors, including 30 RouteViews~\cite{routeviews} and 24 RIPE RIS~\cite{ripe-ris} collectors. While most of these collectors are positioned in North America and Europe, every populated continent is represented by at least one collector. \\

\noindent\textbf{Measurement Limitations:} While our framework allows us to effectively probe the control plane for evidence of Peerlock and related techniques, a number of limitations prevent complete certainty regarding Peerlock filter placement. The most important of these obstacles are imperfect collector coverage, topological instability, and the presence of other filtering systems. Here we discuss each of these factors in turn.

BGP policies prohibit us from viewing the entirety of the topology with our framework; there are few collectors in stub networks, and stub/remote ASes do not export received updates back "up" through provider networks. This means our \textit{observation window} - the ASes on update paths at collectors - is biased toward transit networks in the Internet's core as in~\cite{oliveira2009completeness}. Fortunately, this is the most important/influential region to monitor, as these network's policies have the widest impact on the control plane. Altogether, we observed 610 ASes during our experiments, including 181/605 large ISPs and all 19 Tier 1 networks. Most observed ASes (332) were present in the observation window during all experiments conducted from August 2019-May 2020. Note that while we can only infer \textit{protector} ASes from our observation window, we can poison any AS. So, our window does not limit our inference regarding which ASes are \textit{protected}.

To account for instability in our observation window, we limit our filtering inferences to those ASes observed in control updates both before and after the leak advertisement (i.e., for the current and next target AS experiment). Additionally, we repeat experiments - issue control/leak advertisements for the same target ASes - over several months. These observations are combined to reduce the "noise" of topological dynamism from our inferences. Specifically, we remove ASes from a target's filtering inference sets if we later observe them propagating a leak update for that target; in this case, the earlier inference was likely caused by the AS's intermittent presence in the observation window during the experiment.     

Most importantly, we acknowledge that we cannot be certain ~\pl{}/Peerlock-lite exactly as described by NTT~\cite{peerlock-talk, peerlock} is responsible for all observed filtering, but our experiments are designed to avoid common leak filtering systems. First, since the leak and control advertisements in our experiments share an origin AS/prefix, their updates present identically for ROV filtering purposes. Additionally, since we observe all ASes in the control DAG propagating control updates, we infer those ASes will not apply common IRR or max-prefix limit filtering to the same /24 in leak updates. Finally, while prior work indicates that short poisoned paths are frequently present on the control plane~\cite{tran2019feasibility} and rarely filtered~\cite{smith2020withdrawing}, the poisoning in the control advertisement ensures that we do not conflate poison filtering and \pl{}ing.

Despite our efforts to avoid common filtering practices, local policies grant network operators extensive discretion in how routes are vetted and exported. This flexibility means we cannot be certain that experimental updates are not sometimes blocked by AS specific, ad-hoc AS PATH filtering techniques. We know of no way to distinguish such functionally similar filters from \pl{}. \\

\noindent\textbf{Ethics:}
We issued only well-formed BGP advertisements using the PEERING software client and adhered to all rules published by PEERING. We advertised only our assigned /24 prefixes, which are reserved for experimental use, and thus did not disturb Internet control or data plane operation for any non-experimental IP addresses. Our experiments did require poisoned advertisements, but this is a common practice used both in research~\cite{birge2019sico, smith2020withdrawing} and in traffic engineering~\cite{tran2019feasibility}. One network operator observed and inquired about our experiments to PEERING, but did not report any resultant adverse effects. No data-plane traffic was sent during the conduct of our experiments.
\begin{figure*}[h]
    \centering
    \subfloat[Number of protector/protected rules by ASN. Protector numbers include ASes protecting their own ASN via loop detection.]{\label{fig:t1_bar}\includegraphics[width=0.45\textwidth]{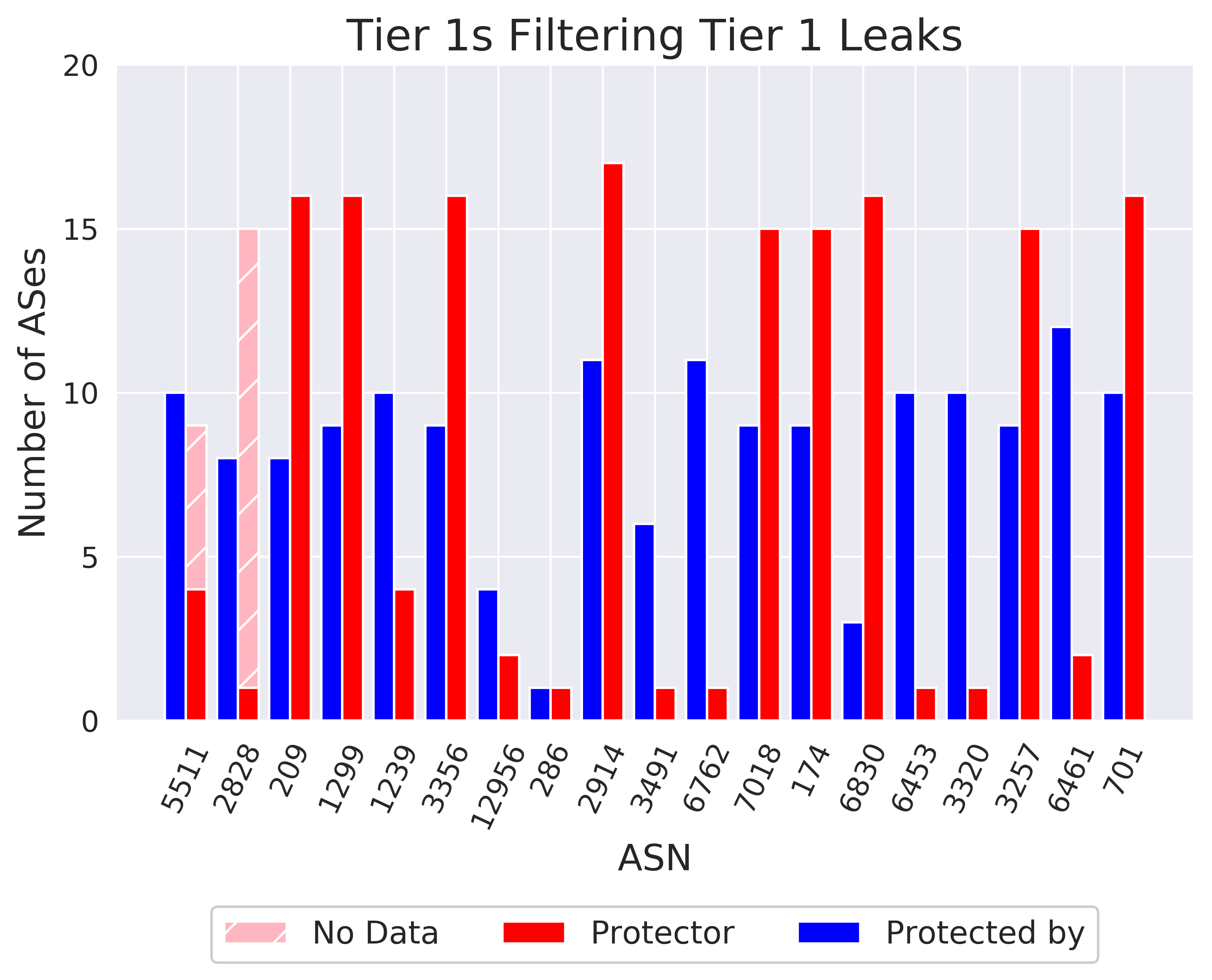}}
    \hspace{0.05cm}
    \subfloat[Depiction of Tier 1 protection rules.]{\label{fig:t1_recip}\includegraphics[width=0.45\textwidth]{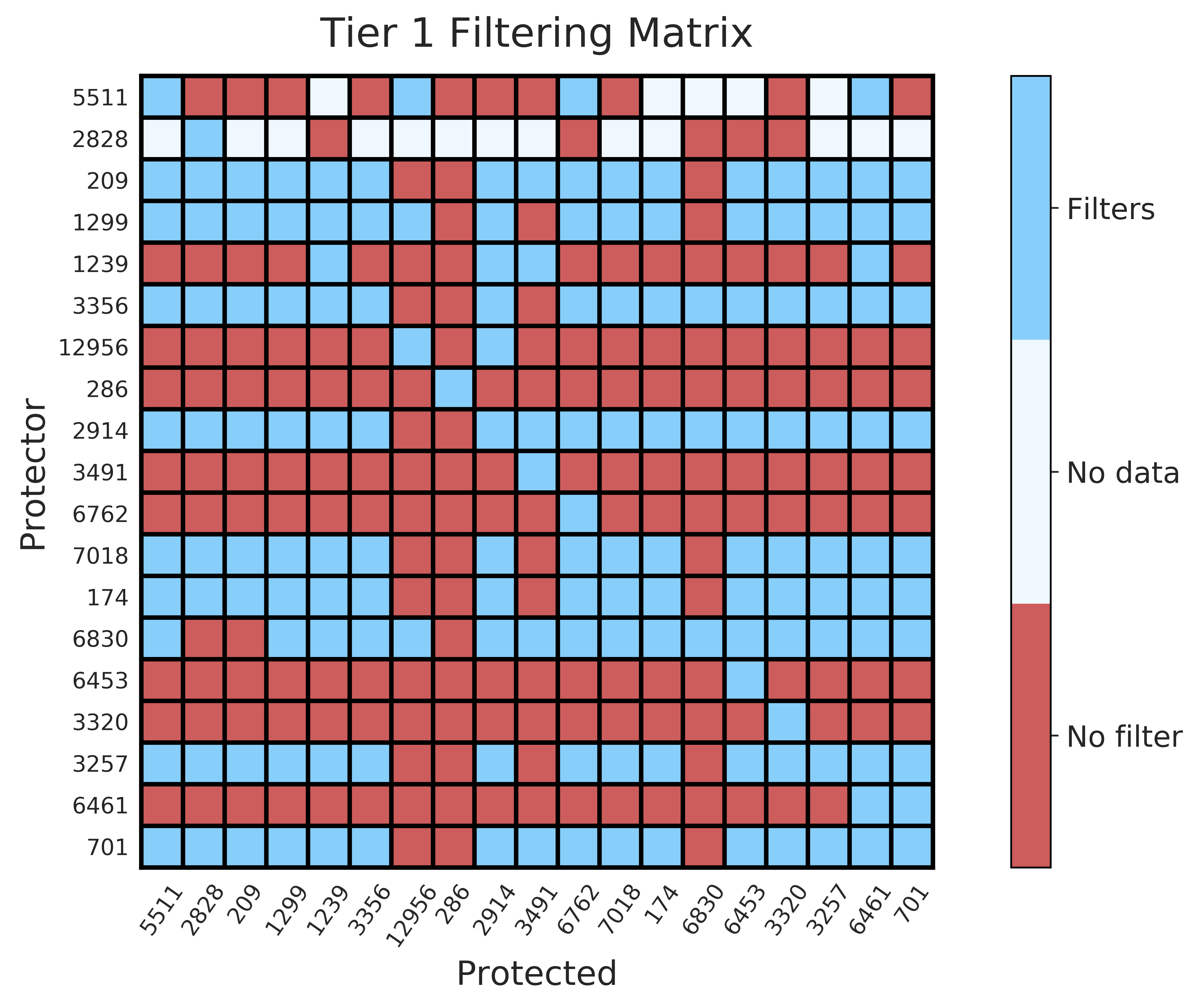}}
    \hspace{0.05cm}
    %\subfloat[Caption here]{\label{fig:link-rel-dist}\includegraphics[width=0.29\textwidth]{graphs/measurement_graphs/t1_long.png}}
    \caption{Tier 1s filtering Tier 1 leaks, 2019/2020 measurements.}
    \label{fig:tier1_measure}
    	\vspace{-15pt}
\end{figure*}

\subsection{Evaluation}{\label{inference}}

\noindent \textbf{Target Set 1, Tier 1s}: 
The 19 Tier 1 ASes form our first target AS set, i.e. the potential protected ASes for which we are inferring \pl{} rules. The Tier 1 peering clique includes the most influential networks by one of the few observable metrics, customer cone size~\cite{CAIDA}, and often creates~\cite{dyn-level3leak} or distributes~\cite{verizon-cloudflareleak, google-leak, rostelecom-leak} leaks that disrupt global Internet services. Paradoxically, deploying filters for leaks that include Tier 1 ASes is also relatively simple for non-Tier 1 networks via the Peerlock-lite system described above. We iteratively issued unpoisoned, control, and leak advertisements that covered this set every two months from August 2019 to May 2020. This repetition allows us to capture filtering rules for ASes with inconsistent presence in our observation window, and to explore how deployments change over time. 

We first present results for protection within the Tier 1 clique in Fig.~\ref{fig:tier1_measure}. Note that because of BGP loop detection, every AS filters leak updates that include their own ASN regardless of \pl{} deployment. The peering clique is fortunately the most stable feature in our observation window, enabling us to measure the presence/absence of nearly every potential \pl{} rule within the clique. We have marked the exceptions for which we were unable to measure filtering rules in pink in Fig.~\ref{fig:tier1_measure}. We see that \pl{} deployment is significant but unevenly distributed within the clique. Some ASes - e.g. AS 2914 NTT, AS 701 Verizon - filter leak updates for virtually the entire clique. For five others - e.g. AS 3491 PCCW Global, AS 6762 Telecom Italia - we found no evidence of Tier 1 \pl{} filtering at all.

\begin{figure*}[t]
    \centering
    \subfloat[Blue lines show poison filtering; red lines depict Tier 1 leak filtering.]{\label{fig:t1_overall}\includegraphics[width=0.45\textwidth]{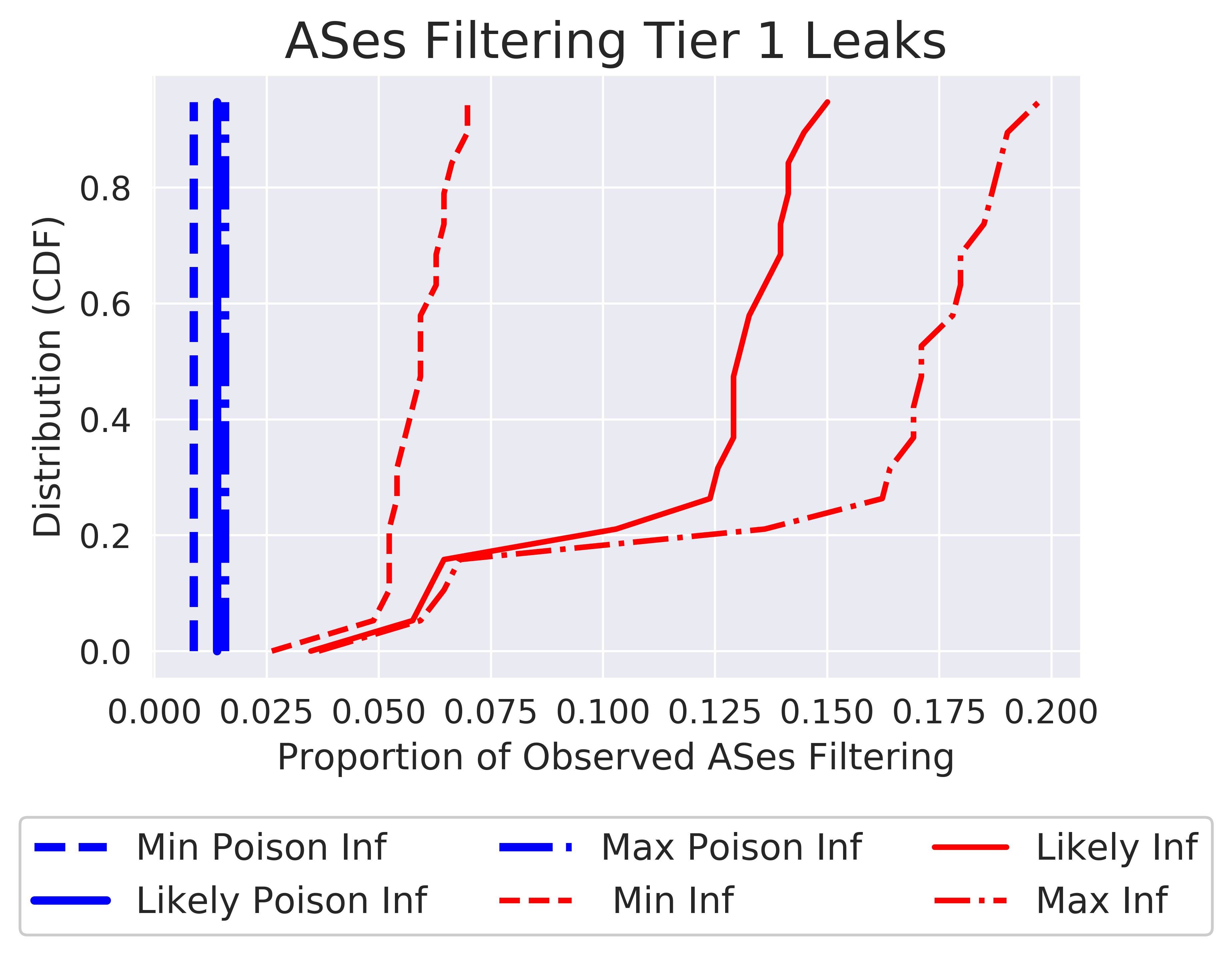}}
    \hspace{0.05cm}
    \subfloat[Blue bars show no. ASes in observation window; red bars show no. ASes filtering at least 1 Tier 1 leak.]{\label{fig:tier_bar}\includegraphics[width=0.45\textwidth]{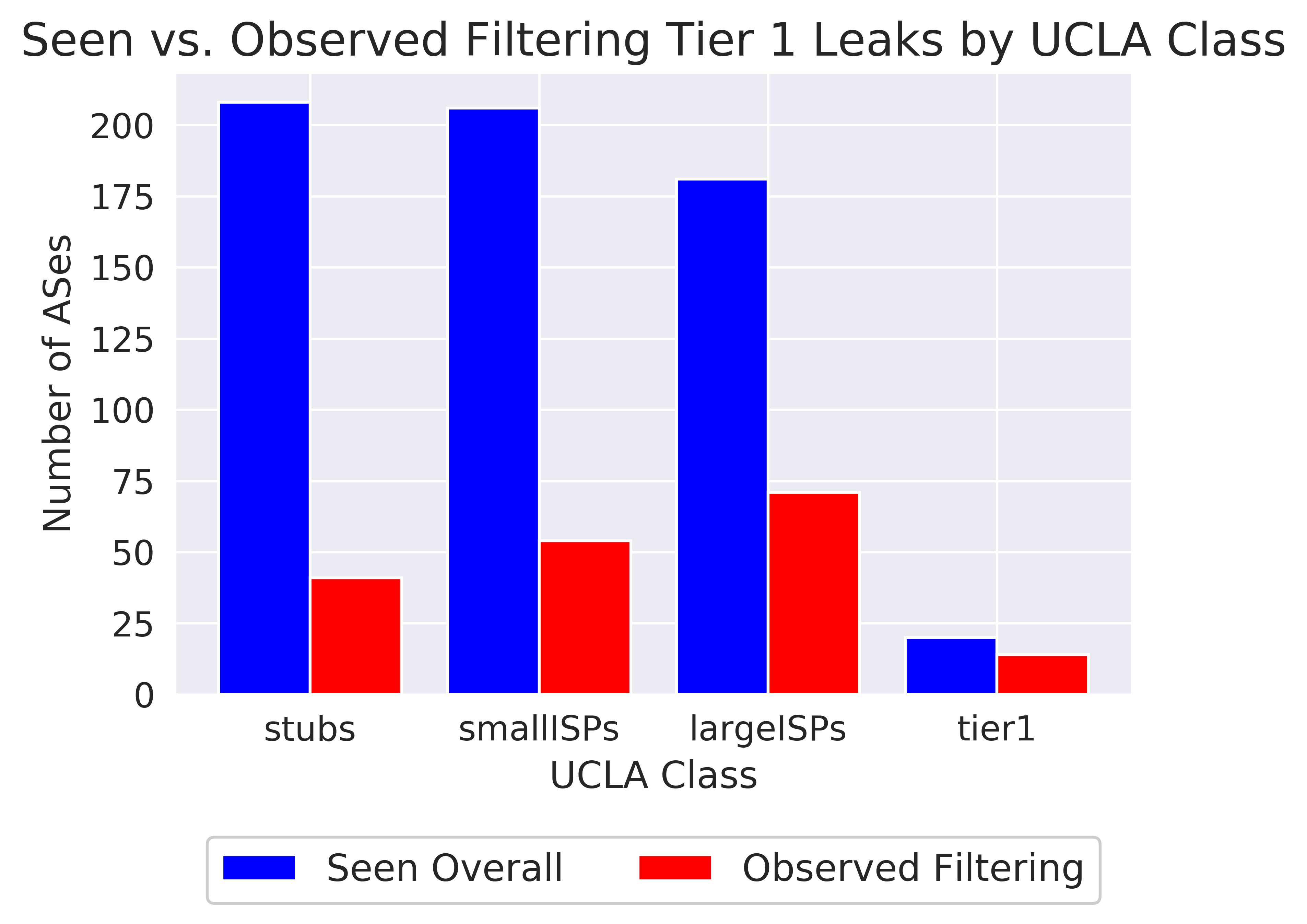}}
    \hspace{0.05cm}
    %\subfloat[Caption here]{\label{fig:link-rel-dist}\includegraphics[width=0.29\textwidth]{graphs/measurement_graphs/t1_long.png}}
    \caption{Overall filtering of Tier 1 leaks, 2019/2020 measurements.}
    \label{fig:tier1_nt1_measure}
\end{figure*}

Our measurement results for \pl{}/Peerlock-lite protection of Tier 1s by all observed ASes are depicted in Fig.~\ref{fig:tier1_nt1_measure}. Fig~\ref{fig:t1_overall} shows both our inferences about which networks filter poisoned updates in general (blue lines) and which filter Tier 1 leaks (red lines). These are displayed as a cumulative distribution function (CDF) over Tier 1 targets; likely inferred filtering levels range from about 3\% (AS 6830) to 15\% (AS 701) of observed ASes. Note that per the experimental design described above, we cannot make \pl{} protection inferences for ASes filtering all poisoned updates; however, this is a small set without Tier 1/large ISP members (max inference size = 9 ASes). Fig.~\ref{fig:tier1_nt1_measure} shows the number of ASes in each UCLA class (see Section~\ref{background}) protecting at least one Tier 1 target. \\

\noindent \textbf{Target Set 2, Tier 1 Peers}: Our second target set includes the non-Tier 1 peers of Tier 1 ASes (about 600 ASes) as inferred by CAIDA~\cite{CAIDA}. These experiments explore whether Tier 1 ASes are extending \pl{} protection to their non-Tier 1 peers. Additionally, despite covering about 1\% of all ASes, this set includes a third of all large ISPs. The presence of these large ISPs in the target set allows us to investigate whether non-Tier 1 ASes apply Peerlock-lite filters to large transit networks outside the peering clique. These experiments were conducted from Oct 2019 to May 2020, with every included network targeted at least twice.

\begin{figure*}[t]
    \centering
    \subfloat[Overall filtering levels for Tier 1 peer leaks. Max and likely poison inferences match for this set.]{\label{fig:nt1_overall}\includegraphics[width=0.45\textwidth]{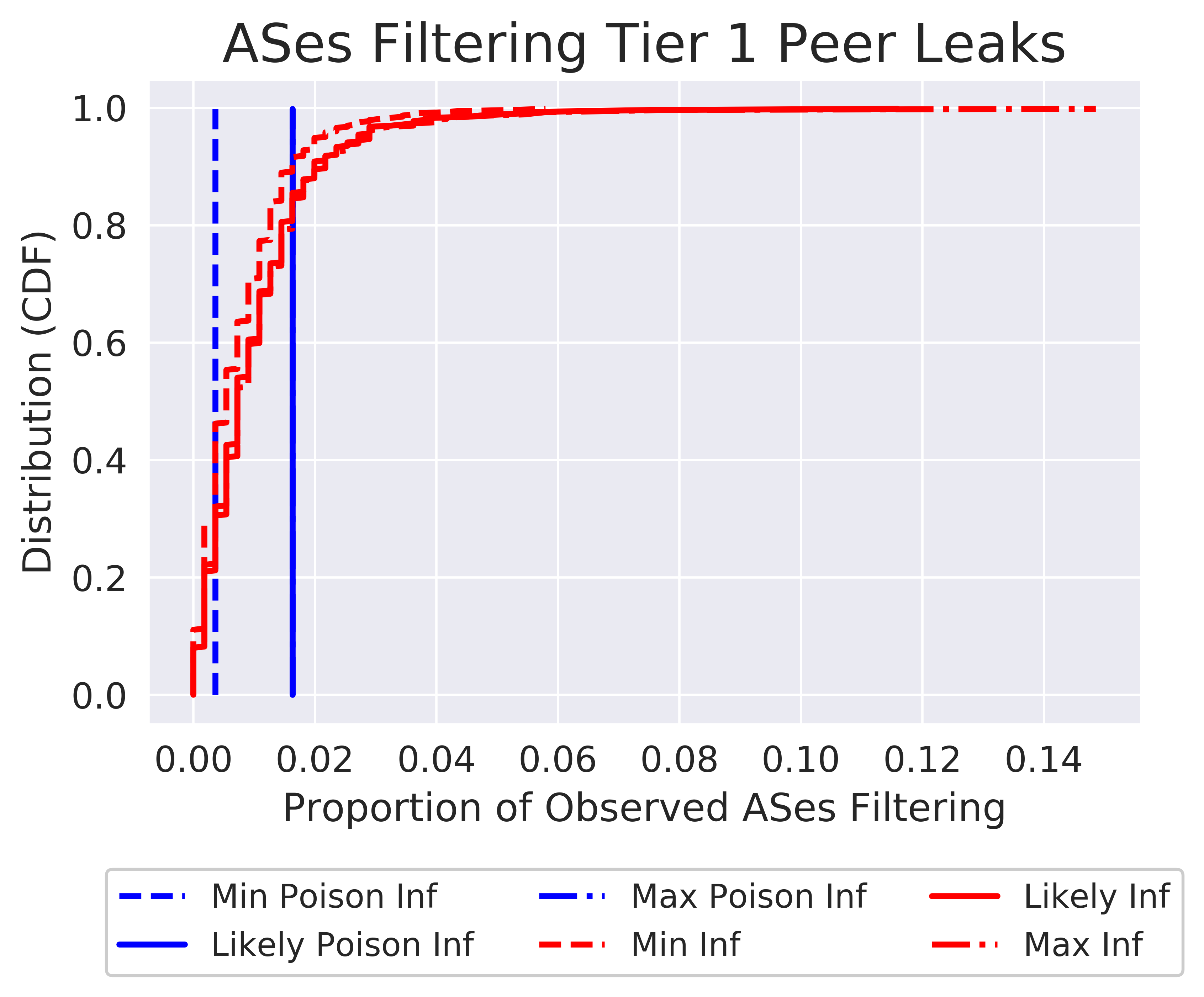}}
    \hspace{0.05cm}
    \subfloat[Tier 1 filtering of Tier 1 peer leaks (peers within clique excluded).]{\label{fig:nt1_bar}\includegraphics[width=0.45\textwidth]{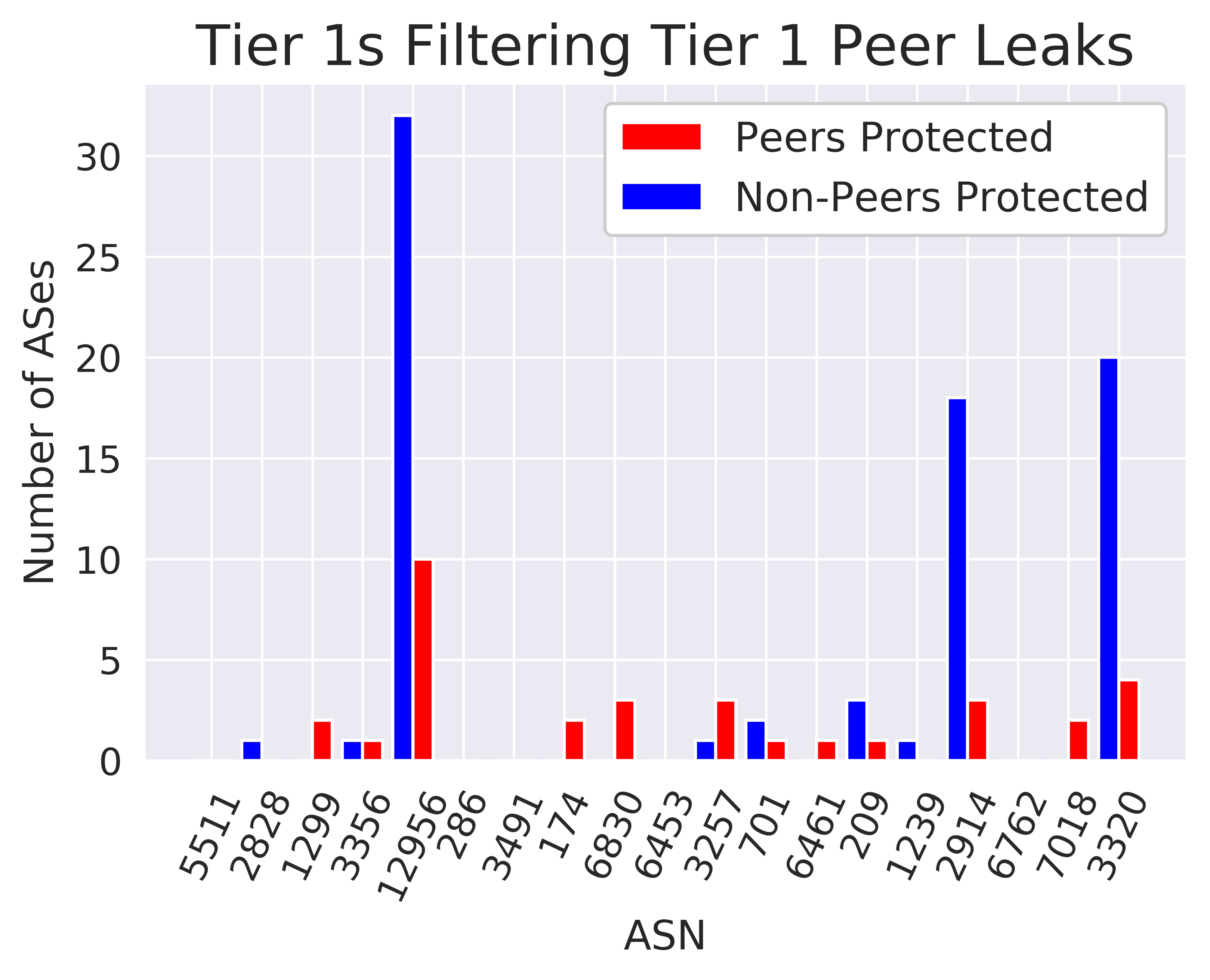}}
    \hspace{0.05cm}
    \caption{Tier 1 peer leaks, 2019/2020 measurements.}
    \label{fig:tier1_peer}
\end{figure*}

The overall results are presented in Fig.~\ref{fig:nt1_overall}. Clearly, filtering for these leaks is less prevalent within our observation window. 80\% of Tier 1 peer leaks were filtered by fewer than 2\% of observed ASes, but a few exceptional targets did trigger significant filtering behavior. Our poison filtering inference for these targets is, as expected, nearly identical to that derived from the Tier 1 leak experiments. Fig.~\ref{fig:nt1_bar} displays filtering levels for each Tier 1 ASes by peering status with the target. All Tier 1s protect 10 or fewer peer networks from this set. More variance exists in non-peer filtering behavior, as we will explore in the following discussion.

\subsection{Discussion}{\label{inference-discussion}}

Consistent with Smith et al.~\cite{smith2020withdrawing}, we find no evidence for widespread filtering of otherwise unremarkable poisoned paths. Their study also found that poisoning high degree ASes in an update is associated with reduced propagation. Specifically, sub-20\% update propagation rates were observed for some Tier 1 ASes, including AS 174 (Cogent/Tier 1) and AS 3356 (Level 3/Tier 1). Birge-Lee et al.~\cite{birge2019sico} likewise found that using AS poisoning rather than communities as a path export control primitive significantly reduced update spread, especially when large transit providers were poisoned. Defensive AS-path filtering (e.g.,\pl{}/Peerlock-lite) is identified as a likely culprit for this effect. Our work systematically examines how and where these filters are deployed on the control plane (within the limits of our observation window). \\

\noindent \textbf{Tier 1 Leak Filtering}:
The greatest protection within our observation window is clearly afforded to Tier 1 ASes. Our initial experiments in August 2019 discovered evidence for 133/342 ($19^2 - 19$) possible Tier 1-Tier 1 filtering rules (about 39\%). Each measurement that followed uncovered at least two new filtering rules, and by our final experiment in May 2020, 153 rules had been observed, a nearly 15\% uptick in \pl{} deployment. We had previously observed a negative filtering result for every additional rule, indicating this increase results from genuinely new \pl{} deployments rather than instability in the observation window. 

Non-Tier 1 ASes also filter Tier 1 leaks, though this behavior is far from uniform. Overall, Tier 1 leak filtering ranged from 3\% to 15\% of observed ASes across Tier 1 AS targets. Most of this is likely due to Peerlock-lite filtering, as it is simpler to deploy. Moreover, fewer than 10\% of the more than 1,000 observed Tier 1 filtering rules exist between peers, and only about 20\% (236 rules) involved a Tier 1's indirect customers filtering leak updates. This suggests that ASes are installing Peerlock-lite filters for all Tier 1s rather than simply protecting their upstream providers. 

\begin{figure}[h]
	\centering
	\includegraphics[width=0.75\columnwidth]{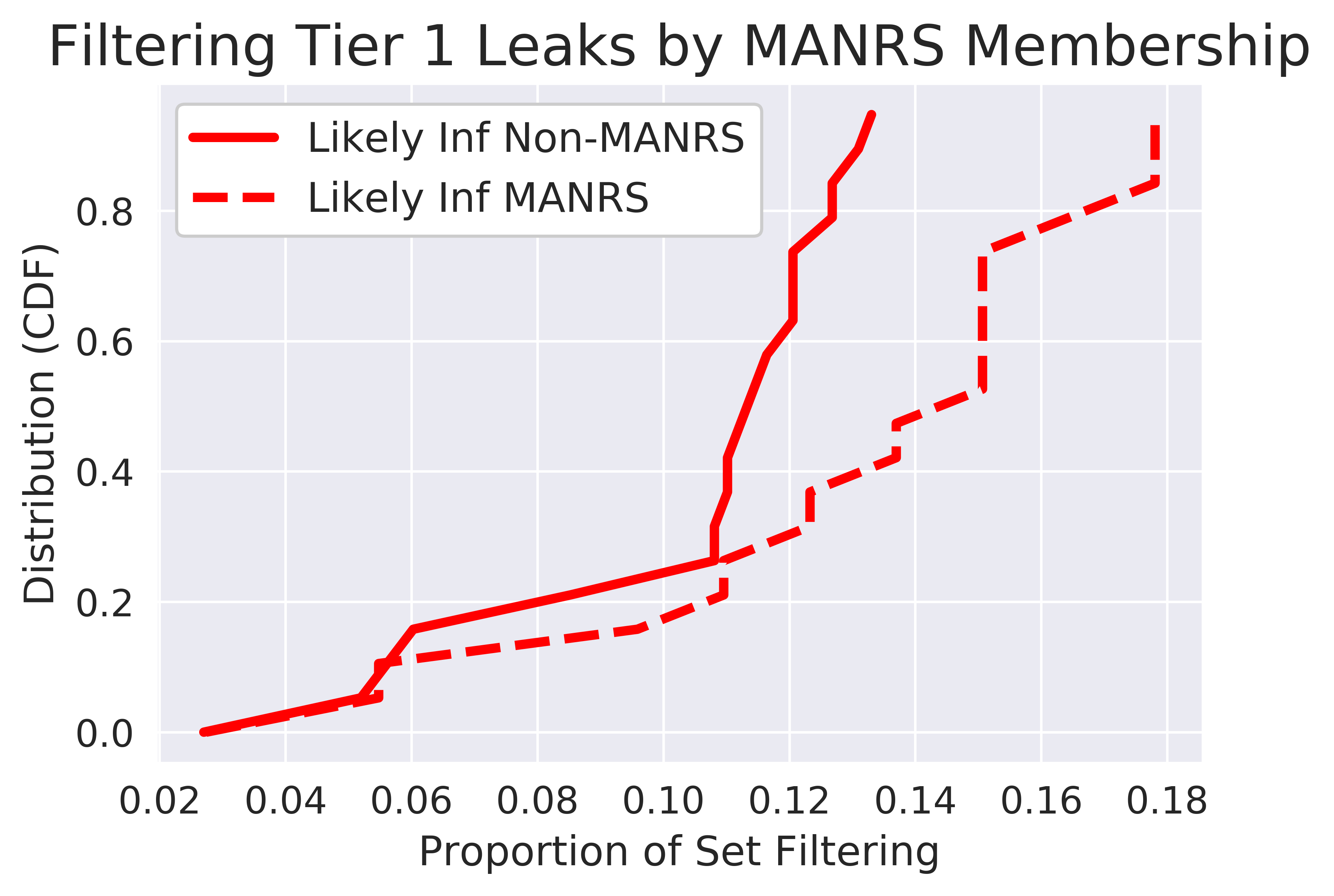}
	\caption{Tier 1 leak filtering for MANRS/non-MANRS ASes.}
	{\label{fig:manrs}}
\end{figure}

Mutually Agreed Norms for Routing Security (MANRS)~\cite{manrs} is an initiative whose ISP members agree to best routing practices (like AS path filtering) to secure inter-domain routing. While \pl{} and Peerlock-lite are not specifically included in MANRS expected filtering actions, they are both suggested in the implementation guide~\cite{manrs-filter}. Fig.\ref{fig:manrs} displays as a CDF the proportion of MANRS and non-MANRS ASes filtering Tier 1 leaks. 73 of 502 MANRS ASes fall within our observation window; the proportion of observed MANRS ASes that filtered Tier 1 leaks ranged from 2-18\% depending on Tier 1 target. Non-MANRS filtering over the same target set ranged from 2 to 12\%. 

As shown in Fig~\ref{fig:tier_bar}, the proportion of ASes with Tier 1 leak filters rises with UCLA class~\cite{oliveira2009completeness}. Intuitively, networks with larger customer cones have the resources for sophisticated configurations and the imperative to prevent issues for downstream customers, and have previously been associated with differing responses to BGP events~\cite{smith2020withdrawing, bush2009internet}. This dynamic hampers systems requiring wide participation like ROV~\cite{gilad2017we} and IRR filtering~\cite{kuerbis2017internet}, but does not limit \pl{} or Peerlock-lite deployment. \\

\noindent \textbf{Tier 1 Peer Leak Filtering}:
Our non-Tier 1 leak experiments met with relatively sporadic filtering. For more than 80\% of targets in this set, nearly every observation window AS (>=98\%) propagated leaks. As described in Section~\ref{overview}, Peerlock-lite filters for non-Tier 1 ASes require more careful deployment. The outliers in this target set (see the long tail in Fig.~\ref{fig:nt1_overall}) are invariably near-Tier 1 networks like AS 1273 Vodafone, AS 6939 Hurricane Electric, and AS 7843 Charter that are safe for most ASes to include in a Peerlock-lite filter. 

Tier 1 filtering of this leak set was likewise reduced compared to Tier 1 leaks. In general, Tier 1 networks deploy fewer than 5 \pl{} filters for non-clique peers. Nearly all of these cover near-Tier 1s like AS 7922 Comcast and AS 1273 Vodafone, or ASes administered by Tier 1s e.g. AS 702/703 Verizon and AS 3549 Level 3. Notably, three networks extend protection to more than 15 non-peers (per CAIDA's inference). AS 2914 NTT's non-peer filtering rules all cover various Comcast ASNs, while AS 12956 Telefonica's rules appear to be regionally-based: zero rules are applied to customer cone ASes, but 23/31 apply to other European ISPs of varying size. 13/20 of AS 3320 Deutsche Telecom's non-peer filtering rules, on the other hand, cover ASes within its customer cone.

In summary, \pl{} is widely deployed and expanding within the peering clique. Deployment outside the peering clique is relatively limited, however. Up to 20\% of non-clique networks also deploy Peerlock-lite (or a similar mechanism) to filter leaks containing Tier 1 or near-Tier 1 ASes. These deployments are proportionally more common in ISPs and rarely seen in stub ASes within our observation window. Fortunately, the effectiveness of \pl{}/Peerlock-lite deployments is less sensitive to scattershot deployment than other filtering solutions. Prior work~\cite{gilad2017we} and our simulations in the following Section~\ref{simulations} suggest that filtering by large ISPs can have an outsize impact on global leak propagation.

\section{Exploring \pl{}'s Practical Impact}{\label{simulations}}The substantial but limited \pl{}/Peerlock-lite filtering measured in the previous section leads us to investigate these systems' protective benefit in partial deployment. We have interest both in how well these systems protect the control plane from Tier 1 leaks as deployed, and in the relative effect of realistic additional deployment (e.g. adding filters at large transit networks). To answer these questions, we quantify \pl{}'s practical impact with Internet-scale leak simulations against several filter deployment schemes. 

\subsection{Simulation Methodology}
These experiments are conducted via extensions to a BGP simulator, an approach consistent with prior work on this topic~\cite{Schuchard:2010io, RAD, smith2018routing, tran2019feasibility}. We construct a simulated AS-level topology from CAIDA's inferred relationship dataset (Jan. 2020 data)~\cite{CAIDA}. ASes within the topology evaluate and export routes using the BGP decision process; longest-prefix matching, LOCAL PREF, and AS PATH guide path selection, while route export is governed by local policy to enforce valley-free routing. This ensures the simulator models the central dynamic of control plane propagation - the Gao-Rexford model~\cite{gao2001inferring}, and allows for the closest approximation of control plane behavior we can devise without ASes' full (private) routing policies.

Each simulation is driven by a \textit{protection scenario} that maps protector ASes to those they are protecting. As with \pl{} in practice, these protectors drop all received routes that transit a protected AS unless they arrive directly from that AS. Some scenarios also include Peerlock-lite deployments; for these experiments, some set of ASes filter all customer-exported routes that transit Tier 1 ASes. Once we establish the protection scenario, we iterate over all Tier 1 to Tier 1 links (with 19 Tier 1 ASes, this is $n = 19, n^2-n = 342$ links). These links describe a unidirectional connection from one Tier 1, called the \textit{link start}, to another Tier 1, called the \textit{link end}. 

\begin{figure}[h]
	\centering
	\includegraphics[width=0.75\columnwidth]{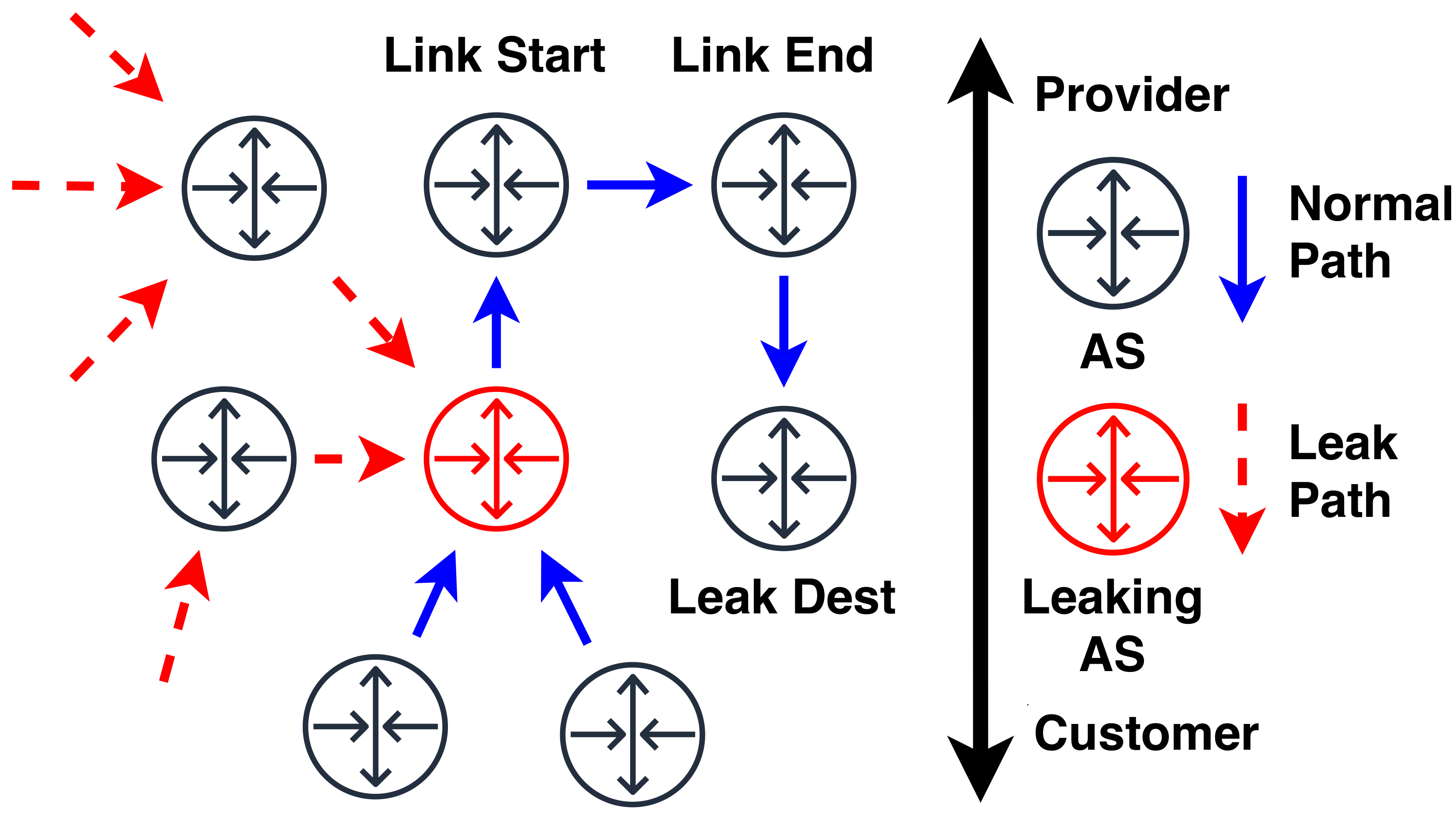}
	\caption{Example simulated leak. Dashed red lines indicate route leak to other providers/peers.}
	{\label{fig:sim_diag}}
\end{figure}

For every link in this set, we sample 20 ASes from the link start's customer cone to serve as \textit{leakers}. Each leaker will, in turn, randomly select a \textit{destination} AS in the link end's customer cone, and advertise a route to the destination over the link to all of its peers/providers (see Fig.~\ref{fig:sim_diag}). This models a Type 1 route leak of a path over the peering clique~\cite{leak-rfc}. After the leak, we allow the topology to converge and measure how many ASes 1) received leak updates and 2) installed the leak path. Additionally, we capture all the AS PATH of all leak updates for analysis. With 20 leaker/destination pairings per link and 342 Tier 1 links, we simulate 6,840 leaks in total. 

Our simulations focus on leaks with Tier 1 leaks for two reasons. First, we do not find substantial real-world \pl{}/Peerlock-lite protection of non-Tier 1 ASes as outlined in Section~\ref{measuring}. Second, many consequential leaks are propagated globally over the Tier 1 backbone, e.g.~\cite{dyn-level3leak, verizon-cloudflareleak, rostelecom-leak, google-leak}. Some of our protection schemes will investigate whether leaks can propagate throughout the Internet without Tier 1 distribution.

\subsection{Evaluation}{\label{sim-eval}}
\begin{figure*}[t]
    \centering
    \subfloat[Impact of various deployment scenarios on leak update propagation.]{\label{fig:sim_heard}\includegraphics[width=0.45\textwidth]{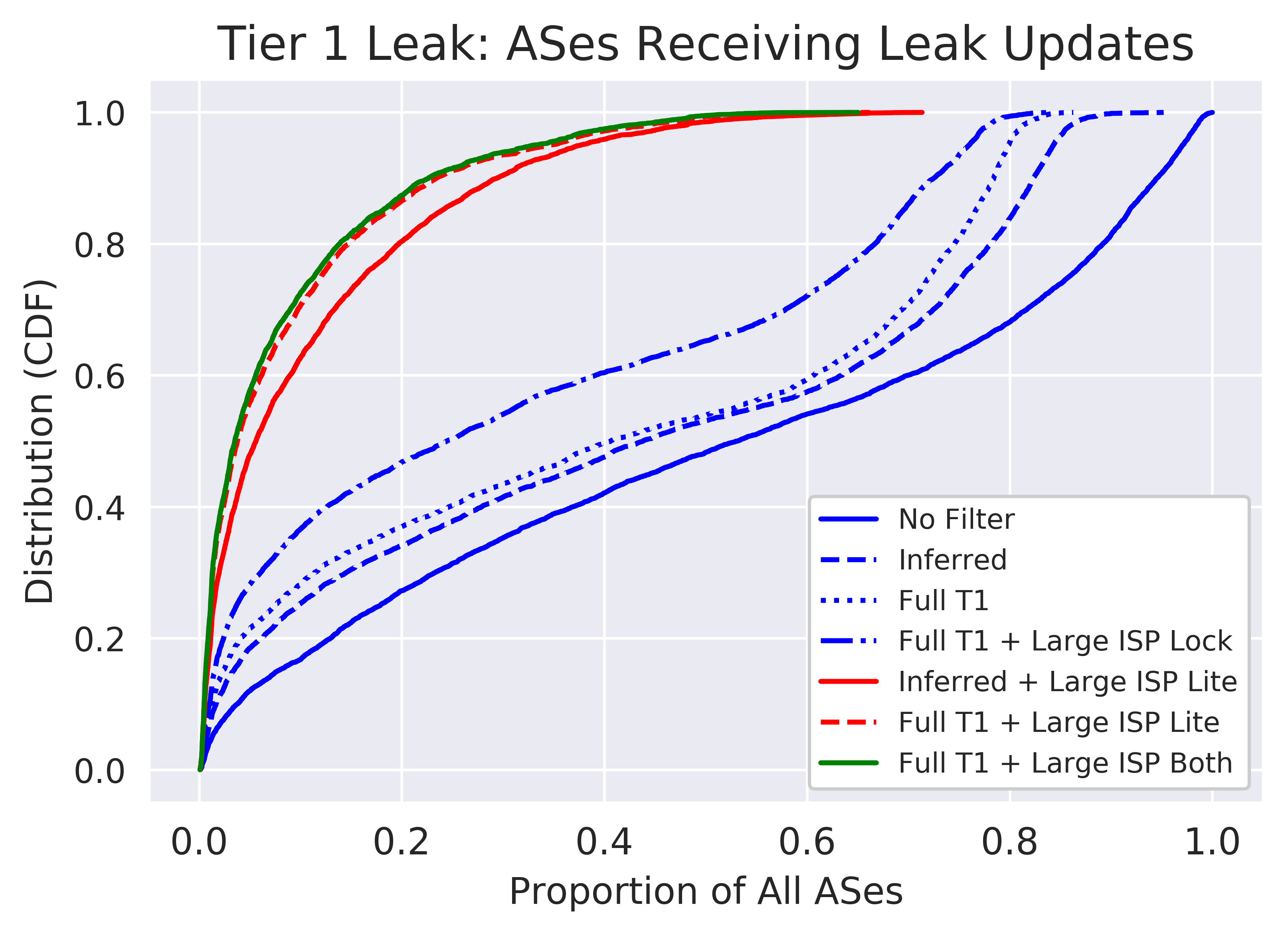}}
    \hspace{0.05cm}
    \subfloat[Note increased Peerlock-lite performance for path switching vs. leak update propagation.]{\label{fig:sim_switch}\includegraphics[width=0.45\textwidth]{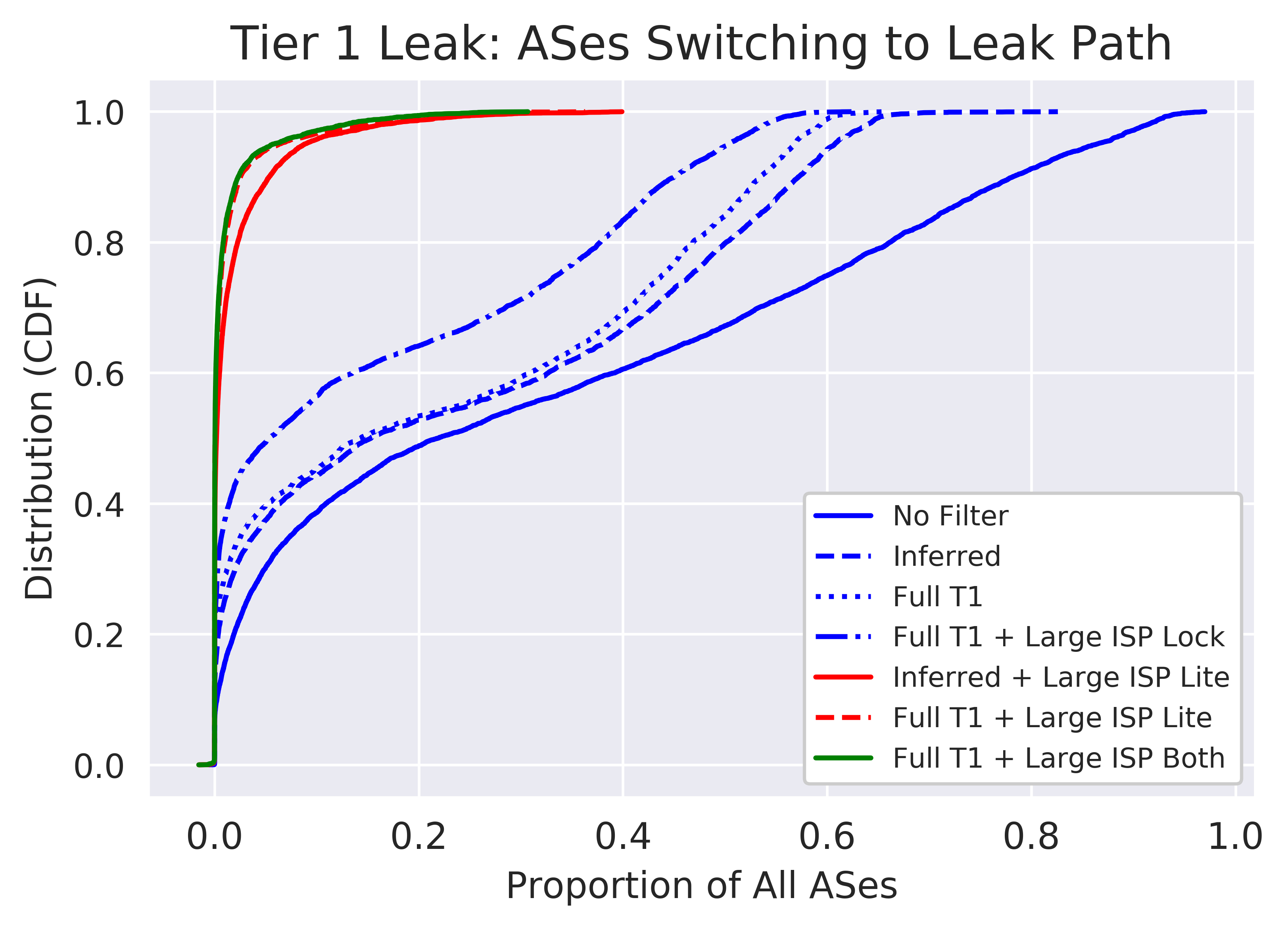}}
    \hspace{0.05cm}
    \caption{Peerlock/Peerlock-lite simulation results.}
    \label{fig:simulator}
    	\vspace{-2pt}
\end{figure*}

We evaluate seven different protection schemes for Tier 1 leaks. 
\begin{itemize}
    \item \textbf{No filters}.
    \item \textbf{Inferred}: Tier 1 \pl{} levels observed during Internet measurements.
    \item \textbf{Full T1}: All Tier 1s \pl{} for all other Tier 1s.
    \item \textbf{Full T1 + large ISP lock}: Same as full T1, but all large ISPs (376 ASes in CAIDA Jan 2020 dataset~\cite{CAIDA}) \pl{} their Tier 1 peers.
    \item \textbf{Full T1 + large ISP lite}: Same as full T1, but all large ISPs deploy Peerlock-lite to protect clique ASes.
    \item \textbf{Full T1 + large ISP both}: Same as full T1, but all large ISPs deploy Peerlock-lite filters and \pl{} for their Tier 1 peers.
    \item \textbf{Inferred + large ISP lite}: Same as inferred, but all large ISPs deploy Peerlock-lite.
\end{itemize}
While it is simpler to filter customer-learned routes with Peerlock-lite than to deploy Tier 1 \pl{} filters for large ISPs, we include both \pl{} and Peerlock-lite filtering by these ASes to study how leaks are propagated within the topology. The results of these experiments are presented in Fig.~\ref{fig:simulator}, which displays both the proportion of ASes in the topology receiving leak updates (Fig.~\ref{fig:sim_heard}), and the proportion selecting/exporting the leak path (Fig.~\ref{fig:sim_switch}).

A critical feature revealed by Fig.~\ref{fig:simulator} is the insufficiency of Tier 1 protection alone (blue lines). Full Tier 1 \pl{}ing prevents all distribution of studied leaks over the peering clique, but leak updates still spread to the majority of the topology for most experiments. Adding large ISP \pl{} protection has a relatively significant impact on both propagation and installation. 

Peerlock-lite deployment by these ASes (red lines) benefits from more filterers with wider protection per filterer. Naturally, these scenarios are much more effective at preventing propagation. For most leak cases, less than 10\% of the topology receives leak updates. This highlights the leverage large ISPs have within the topology; filtering at these ASes (<1\% of all networks) generates an extensive shielding effect. The distinct "shoulder" on the Peerlock-lite curves in Fig.~\ref{fig:sim_switch} suggests the impact on ASes \textit{using} the leak is even more pronounced. There is virtually no impact on target link usage for 75\% of simulated leaks when Peerlock-lite is deployed by all large ISPs. Interestingly, the combination of \pl{} and Peerlock-lite filtering by large ISPs (green line) adds little value over Peerlock-lite alone.

\subsection{Discussion}{\label{sim-discussion}}

\noindent\textbf{Path Encoding:}
To analyze how each of these scenarios shapes leak propagation (and route selection/export), we collect the AS PATH of all leaks exported during the above experiments. We use a novel \textit{path encoding} whereby each AS on leak AS PATHs is converted to a 2-tuple with the form (relationship to next AS, UCLA class~\cite{oliveira2009completeness}). Only the AS PATH segment from the first customer to provider link to the leaker ASN - the \textit{leak segment} - is encoded. This trimming discards the "down" segment prepended as leaks propagate within customer cones, as well as the the segment connecting leaker and destination that is invariant across leaks. We include AS relationship in the encoding because of its importance in path export behavior as described in Section~\ref{bgp}; UCLA class informs us regarding where leaks travel through the routing hierarchy. Taken together, these factors help us understand broadly the topological dynamics at play in leak propagation, and to capture the dominant leak propagation vectors under each protection scenario.

Relationship is encoded as "C" (customer), "R" (peer), or "P" (provider). UCLA classes are indicated by "T" (Tier 1), "L" (large ISP), "S" (small ISP), and "U" (stub). Example: [LR, TP] encodes a leak path exported to a Tier 1 provider by the leaker, who then passes the leak to a large ISP peer. The progress of the leak through the large ISP's customer cone would continue to the left of "LR", and the path from leaker to destination would continue to the right of "TP", but these segments are omitted as explained above.

\begin{table}[]
\begin{center}
\resizebox{\columnwidth}{!}{%
\begin{tabular}{@{}lcc@{}}
\toprule
\multicolumn{3}{c}{\textbf{Common Leak Segment Encodings}}                                                                               \\ \midrule
\textbf{Scenario/Encoding}         & \multicolumn{1}{l}{\textbf{No. exporting ASes}} & \multicolumn{1}{l}{\textbf{\% of exporting ASes}} \\
\rowcolor[HTML]{C0C0C0} 
\textbf{No filters}                & 141,797,992                                     & 100\%                                             \\
{[}LR, LP{]}                       & 14,892,311                                      & 11\%                                              \\
{[}TP{]}                           & 10,254,707                                      & 7\%                                               \\
{[}LR, LP, LP{]}                   & 8,683,968                                       & 6\%                                               \\
\rowcolor[HTML]{C0C0C0} 
\textbf{Inferred}                  & 108,030,704                                     & 100\%                                             \\
{[}LP, LP{]}                       & 14,325,960                                      & 13\%                                              \\
{[}LR, LP, LP{]}                   & 8,675,841                                       & 8\%                                               \\
{[}SR, LP{]}                       & 5,169,427                                       & 5\%                                               \\
\rowcolor[HTML]{C0C0C0} 
\textbf{Full T1}                   & 101,024,444                                     & 100\%                                             \\
{[}LR, LP{]}                       & 14,246,024                                      & 14\%                                              \\
{[}LR, LP, LP{]}                   & 8,978,175                                       & 9\%                                               \\
{[}SR, LP, LP{]}                   & 5,163,786                                       & 5\%                                               \\
\rowcolor[HTML]{C0C0C0} 
\textbf{Full T1 + large ISP lock}  & 69,638,282                                      & 100\%                                             \\
{[}LR, LP{]}                       & 9,473,820                                       & 14\%                                              \\
{[}LR, LP, LP{]}                   & 5,899,779                                       & 8\%                                               \\
{[}SR, LP{]}                       & 3,310,842                                       & 5\%                                               \\
\rowcolor[HTML]{C0C0C0} 
\textbf{Inferred + large ISP lite} & 8,005,724                                       & 100\%                                             \\
{[}LR{]}                           & 2,537,276                                       & 32\%                                              \\
{[}LR, TP{]}                       & 1,281,620                                       & 16\%                                              \\
{[}LR, SP{]}                       & 653,167                                         & 8\%                                               \\
\rowcolor[HTML]{C0C0C0} 
\textbf{Full T1 + large ISP lite}  & 5,215,232                                       & 100\%                                             \\
{[}LR{]}                           & 2,386,597                                       & 46\%                                              \\
{[}LR, SP{]}                       & 679,076                                         & 13\%                                              \\
{[}SR{]}                           & 412,399                                         & 8\%                                               \\
\rowcolor[HTML]{C0C0C0} 
\textbf{Full T1 + large ISP both}  & 4,649,828                                       & 100\%                                             \\
{[}LR{]}                           & 2,023,579                                       & 44\%                                              \\
{[}LR, SP{]}                       & 584,124                                         & 13\%                                              \\
{[}SR{]}                           & 407,661                                         & 9\%                                               \\ \bottomrule
\end{tabular}
}
\\[12pt]
\end{center}
\normalsize{Table 2: Most common encodings with number and percentage of ASes exporting leaks.}
\end{table}\label{table:encoding}

\begin{table*}[!ht]
\begin{center}
\begin{tabular}{@{}lccccccccccc@{}}
\toprule
\multicolumn{12}{c}{\textbf{Transited AS Statistics}}                                                                                                                                                                                                                                                                                                                                                                                         \\ \midrule
\textbf{Scenario}                                                              & \multicolumn{2}{c}{\textbf{Segment Length}}                                       & \multicolumn{3}{c}{\textbf{Tier 1s}}                                                                  & \multicolumn{3}{c}{\textbf{Large ISPs}}                                                               & \multicolumn{3}{c}{\textbf{Small ISPs}}                  \\
\rowcolor[HTML]{C0C0C0} 
\multicolumn{1}{l|}{\cellcolor[HTML]{C0C0C0}}                                  & \textbf{average} & \multicolumn{1}{c|}{\cellcolor[HTML]{C0C0C0}\textbf{std. dev}} & \textbf{\% paths} & \textbf{average} & \multicolumn{1}{c|}{\cellcolor[HTML]{C0C0C0}\textbf{std. dev}} & \textbf{\% paths} & \textbf{average} & \multicolumn{1}{c|}{\cellcolor[HTML]{C0C0C0}\textbf{std. dev}} & \textbf{\% paths} & \textbf{average} & \textbf{std. dev} \\
\multicolumn{1}{l|}{\textbf{No filters}}                                       & 4.4              & \multicolumn{1}{c|}{1.8}                                       & 27\%              & 0.2              & \multicolumn{1}{c|}{0.4}                                       & 89\%              & 2.3              & \multicolumn{1}{c|}{1.6}                                       & 40\%              & 0.7              & 0.8               \\
\rowcolor[HTML]{C0C0C0} 
\multicolumn{1}{l|}{\cellcolor[HTML]{C0C0C0}\textbf{Inferred}}                 & 4.6              & \multicolumn{1}{c|}{\cellcolor[HTML]{C0C0C0}1.8}               & 7\%               & 0.5              & \multicolumn{1}{c|}{\cellcolor[HTML]{C0C0C0}0.2}               & 98\%              & 2.5              & \multicolumn{1}{c|}{\cellcolor[HTML]{C0C0C0}1.5}               & 45\%              & 0.8              & 0.9               \\
\multicolumn{1}{l|}{\textbf{Full T1}}                                          & 4.7              & \multicolumn{1}{c|}{1.8}                                       & 0\%               & 0.0              & \multicolumn{1}{c|}{0.0}                                       & 98\%              & 2.6              & \multicolumn{1}{c|}{1.5}                                       & 46\%              & 0.8              & 0.9               \\
\rowcolor[HTML]{C0C0C0} 
\multicolumn{1}{l|}{\cellcolor[HTML]{C0C0C0}\textbf{Full T1 + large ISP lock}} & 4.8              & \multicolumn{1}{c|}{\cellcolor[HTML]{C0C0C0}2.1}               & 0\%               & 0.0              & \multicolumn{1}{c|}{\cellcolor[HTML]{C0C0C0}0.0}               & 98\%              & 2.7              & \multicolumn{1}{c|}{\cellcolor[HTML]{C0C0C0}1.6}               & 50\%              & 0.9              & 1.0               \\
\multicolumn{1}{l|}{\textbf{Inferred + large ISP lite}}                        & 2.9              & \multicolumn{1}{c|}{1.0}                                       & 35\%              & 0.3              & \multicolumn{1}{c|}{0.5}                                       & 72\%              & 0.6              & \multicolumn{1}{c|}{0.5}                                       & 42\%              & 0.8              & 1.0               \\
\rowcolor[HTML]{C0C0C0} 
\multicolumn{1}{l|}{\cellcolor[HTML]{C0C0C0}\textbf{Full T1 + large ISP lite}} & 2.7              & \multicolumn{1}{c|}{\cellcolor[HTML]{C0C0C0}1.1}               & 0\%               & 0.0              & \multicolumn{1}{c|}{\cellcolor[HTML]{C0C0C0}0.0}               & 74\%              & 0.6              & \multicolumn{1}{c|}{\cellcolor[HTML]{C0C0C0}0.5}               & 47\%              & 1.0              & 1.1               \\
\rowcolor[HTML]{FFFFFF} 
\multicolumn{1}{l|}{\cellcolor[HTML]{FFFFFF}\textbf{Full T1 + large ISP both}} & 2.7              & \multicolumn{1}{c|}{\cellcolor[HTML]{FFFFFF}1.1}               & 0\%               & 0.0              & \multicolumn{1}{c|}{\cellcolor[HTML]{FFFFFF}0.0}               & 71\%              & 0.6              & \multicolumn{1}{c|}{\cellcolor[HTML]{FFFFFF}0.5}               & 49\%              & 1.0              & 1.1               \\ \bottomrule
\end{tabular}
\\[12pt]
\end{center}
\normalsize{Table 3: Analyzing leak segments by UCLA classes transited.}
\end{table*}\label{table:segment}

We will use two tables in analyzing our results. Table 1 depicts the three most common leak encodings for each scenario; these account for at least a quarter of leak paths regardless of filter placement. We also list the sum and percentage of ASes exporting leaks accounted for by each encoding. Table 2 gives summary statistics for leak segments, including their average length and the percentage of leak segments transiting each UCLA class. Because we do not encode customer cone propagation in leak segments, stubs are transited in <10\% of paths across all protection scenarios, are are omitted from the table.

First, we observe that even under the "no filters" scenario, leaks re-transiting the Tier 1 clique are not the most common path encoding in Table 1. Table 2 shows they are present in <35\% of leak segments under all scenarios. This result is an artifact of the BGP decision process; paths learned from customers are preferred over those exported by peers, and peer routes are favored over provider-learned ones. So, with all other selection criteria equal, routes exported from providers "above" an AS in the topology - e.g. the peering clique - will generally only be installed and exported if the AS has not received an update from peers/customers "below". Since Tier 1 providers cap the routing hierarchy, we expect ASes will prefer non-Tier 1 routes when provided alternatives by their connectivity. This dynamic explains why the additional protection afforded by complete \pl{} within the peering clique vs. current levels is muted in Fig.~\ref{fig:sim_switch}.

This effect also brings large ISPs to the fore in our simulations. As noted in~\cite{oliveira2009completeness}, these networks are densely connected with peering links. Their connectivity allows them to bypass the Tier 1 clique for many routes - and makes them the primary channel for leak propagation. The most common encoding for every scenario in Table 1 includes a large ISP, and 18/21 of the top encodings transit at least one. More than 70\% of leak segments transit these ASes for all protection scenarios (see Table 2). In fact, in the scenarios without Peerlock-lite (top four listed), leak segments on average transit - and could be filtered by - multiple large ISPs. These statistics motivate the scenarios that place Peerlock-lite filtering at these ASes (bottom three in tables).

Interestingly, Peerlock-lite diminishes leak usage and propagation unequally as shown in Fig.~\ref{fig:simulator}. Fig.~\ref{fig:sim_heard} shows about 20\% of leak segments propagate to 20\% or more of the topology with large ISP Peerlock-lite deployment, but Fig.~\ref{fig:sim_switch} shows that fewer than 5\% are installed/exported by at least 20\% of ASes. Table 1 hints at why this is the case - a third or more of leak segments in Peerlock-lite scenarios are exported to large ISP peers, who propagate them directly into their customer cones (indicated by [LR]). Large ISPs with any customer-learned or preferential (e.g. shorter) peer-learned paths to the leak destination will prefer their existing route, so the [LR] only includes a subset of the leaker's peers. Large ISP peers advertising the leak to customers could reach many ASes, but as a provider-learned route, the leak will be disadvantaged in the BGP decision process. 

We see in Table 1 and Fig.~\ref{fig:simulator} that small ISPs do not have the connectivity to propagate leaks globally when the large ISP provider channel is blocked by Peerlock-lite. Under all scenarios, most leak segments do not transit a small ISP (though they may be transited during propagation into customer cones). This feature suggests a less prominent role in route exchange for these networks relative to large ISPs. 

To summarize, we find large ISPs are the most critical players in halting the spread and installation of Tier 1 leaks. These networks are interconnected enough to globally disseminate route leaks \textit{without the peering clique} in many cases. Moreover, adding simple Peerlock-lite filters at these ASes to the currently deployed \pl{} filters in the peering clique causes a 94\% reduction in total leak export across 6,840 leak simulations. Table 1 suggests that peer connections among ISPs are the largest remaining vulnerability for Tier 1 leaks given uniform large ISP Peerlock-lite deployment. These channels are out of reach for Peerlock-lite as described, but could be mitigated by 1) additional peering relationships/\pl{} rules to protect important leak targets and/or 2) complementary leak prevention systems like IRR filtering.

\section{Related Work}{\label{related}}Smith et al.'s 2020 study on the efficacy of poison filtering for inbound re-routing~\cite{smith2020withdrawing} similarly employed the PEERING framework to probe the behavior of remote networks. That work encountered some evidence for poison filtering, and noted that filtering rates increase with poisoned AS degree, but did not seek to describe the underlying filtering mechanism or measure which ASes filter poisons. Similarly, Birge-Lee et al.~\cite{birge2019sico} and McDaniel et al.~\cite{mcdaniel2020maestro} trialed poisons as primitives for novel BGP attacks. Both studies encountered filtering when attempting to poison large transit networks, but did not examine filtering position or prevalence. 

Hlavacek et al.~\cite{hlavacek2019disco} introduced the DISCO system for preventing BGP hijacking. While not designed to prevent route leaks, the approach taken by DISCO is ideologically similar to \pl{} - DISCO emphasizes deployability/usability at the expense of some security guarantees relative to RPKI/ROV filtering. This line of thinking is informed by a long history of glacial deployment rates for security features that harden BGP including BGPSec~\cite{sriram2019resilient} and the RPKI~\cite{gilad2017we, rpkimon}.   

Previous work that relies on BGP poisoning often assumes 1) unpoisoned ASes will forward poisoned updates and 2) poisoned ASes will drop such updates (see Section~\ref{poisoning}). Katz-Bassett et al.'s LIFEGUARD fault detection and remediation system~\cite{katz2011machiavellian, katz2012lifeguard}, for instance, employs poisoning to steer ASes around link failures. Smith and Schuchard's Nyx defense~\cite{smith2018routing} depends on rerouting with poisons for DDOS/Link Flooding Attack mitigation. Anwar et al.'s path discovery technique~\cite{anwar2015investigating} is also driven by BGP poisoning. While we discovered little evidence for general poison filtering, the prevalence of Tier 1/large transit network filtering could present an obstacle to these systems. Specifically, the assumption that unpoisoned networks will propagate poisons does not hold in all cases.

\section{Conclusion}{\label{conclusion}}This work probes the current deployment of \pl{}/Peerlock-lite on the control plane with active Internet measurements in Section~\ref{measuring}. We find substantial evidence for deployment of these leak defense systems, especially in large transit networks, and measure a rise in \pl{} deployment within the peering clique during our experiments. While the range of protected networks is still narrow within our observation window, with most filterers protecting only Tier 1 ASes, many of the most disruptive recent route leaks contain these networks. Defensive systems~\cite{smith2018routing, katz2012lifeguard}, measurement techniques~\cite{anwar2015investigating}, and attacks~\cite{birge2019sico} that may poison \pl{}-protected networks could inadvertently trigger these filters, and must not assume poisons will be propagated by unpoisoned networks. BGP simulators should likewise account for the presence of \pl{} to faithfully reproduce control plane behavior.  

We also examine how the position and prevalence of filtering impacts leak propagation in the AS-level topology in Section~\ref{simulations}. Notably, we find that large ISPs filtering plays a major role in global leak dissemination, signaling that Tier 1 clique deployment of \pl{} alone is not sufficient to isolate leaks. Strategic placement of filters at these large transit providers, which account for fewer than 1\% of all ASes, completely mitigates 80\% of simulated Tier 1 route leaks. 

The MANRS filtering guide encourages AS PATH filtering by member ISPs, particularly for screening customer advertisements, and gives \pl{}/Peerlock-lite as examples. But these systems are not explicitly required, unlike IRR filtering (see~\cite{manrs-filter} Section 4.1.1.1). Given the many indirect/direct customers these networks serve, ISPs are best equipped and best incentivized to deploy effective filters. Moreover, neither \pl{} nor Peerlock-lite is technically complex or burdensome to configure. Therefore, we argue for broad application of these common-sense leak prevention techniques by ISPs as a meaningful step in securing inter-domain routing.

\subsection{Operator Engagement}

This study's preprint was posted to the NANOG and RIPE operator mailing lists in June and August 2020. While operator response was limited, email correspondence around the study did yield some helpful insights. One European AS operator claimed that at least one (and possibly additional) transit networks deployed techniques similar to Peerlock around 2007, roughly ten years before NTT's codification of the method~\cite{peerlock}. Separately, a Tier 1 network operator suggested that 1) differences in network automation sophistication could account for observed uneven filtering within the peering clique and 2) defensive filtering may have partially mitigated the Verizon/Cloudflare leaks detailed in Section~\ref{leak-bg}. This idea is supported by Cloudflare's discussion on the incident~\cite{verizon-cloudflareleak}. Cloudflare identifies some networks that filtered the leaks (including ASes 1299 Telia, 2914 NTT, and 7018 AT\&T). Bandwidth graphs presented in that post indicate little or no impact on Cloudflare-to-AT\&T data plane operation, while Cloudflare traffic to leak propagator Verizon was drastically reduced for hours after the incident.

\subsection{Future Directions}

Widespread adoption of \pl{} will likely depend on addressing scalability issues. Rule configuration currently requires non-standard, manual out-of-band communication between protector/protected ASes. Automating this process is a crucial step in extending \pl{} beyond core networks. Communities designating authorized upstreams for routes, as proposed in~\cite{automatepl-draft}, could take the place of out-of-band communication. Alternatively, RPKI registration of direct/indirect customers~\cite{automatepl-draft-cones} could distribute trusted topological information relevant to filtering. 

\section*{Acknowledgements}
The authors would like to thank Samuel Jero for his helpful guidance in the shepherding process. The detailed and thoughtful comments/criticisms from our NDSS reviewers are also appreciated, especially within the context of the ongoing COVID-19 pandemic. Finally, we would like to recognize Job Snijders at NTT Communications for presenting the \pl{} system. This study was supported by the National Science Foundation under Grant No. 1850379.

\bibliographystyle{IEEEtranS}
\bibliography{refs}

\appendix
\noindent\textbf{Update Propagation:}{\label{propagate}}
Before (August 2019) and after (May 2020) conducting our control-plane experiments in Section~\ref{measuring}, we performed simple tests to measure 1) the time distribution for BGP update arrivals at RIPE/RouteViews collectors for normal and poisoned advertisements issued from PEERING, and 2) the time distribution for unique ASes seen on AS PATHs in those updates. The latter is most critical for our experiments, as we build our filtering inferences from the presence/absence of ASes on update AS PATHs.

These tests consisted of an explicit /24 withdrawal followed by a one hour waiting period, then a normal /24 advertisement. We listened for updates for the /24 at all BGPStream collectors, and recorded the arrival times of updates for the advertised prefix for one hour. We also noted when unique ASes were first seen on the updates' AS PATHs. This process (withdraw, update, listen) was repeated five times. We conducted the same process with a poisoned /24 advertisement, for a total of 10 advertisements per experiment.

The results are shown below, Figs.\ref{fig:propo} and~\ref{fig:unique}. About 80\% of updates triggered by a normal or poisoned /24 advertisement that arrived within an hour were received within 30 minutes post-origination in the August experiment. In May, more than 95\% of updates fell within this period. More importantly, for every experiment, all unique ASes seen on update paths over the hour listening window arrived within the first 25 minutes post-origination. Over 95\% of unique ASes were seen within 7 minutes post-origination.

\begin{figure*}[h]
    \centering
    \subfloat[August results.]{\label{fig:propo_aug}\includegraphics[width=0.45\textwidth]{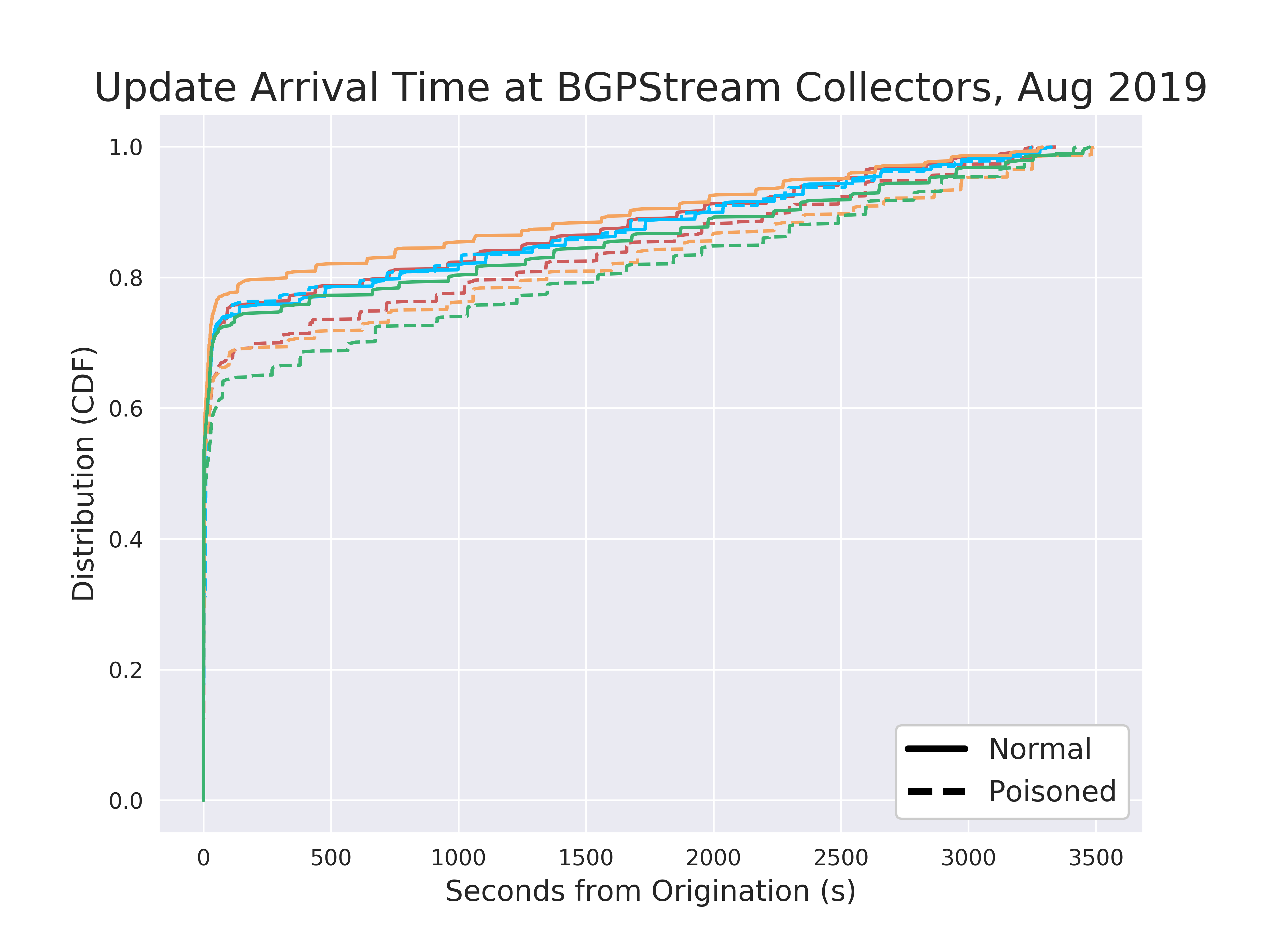}}
    \hspace{0.05cm}
    \subfloat[May results.]{\label{fig:propo_may}\includegraphics[width=0.45\textwidth]{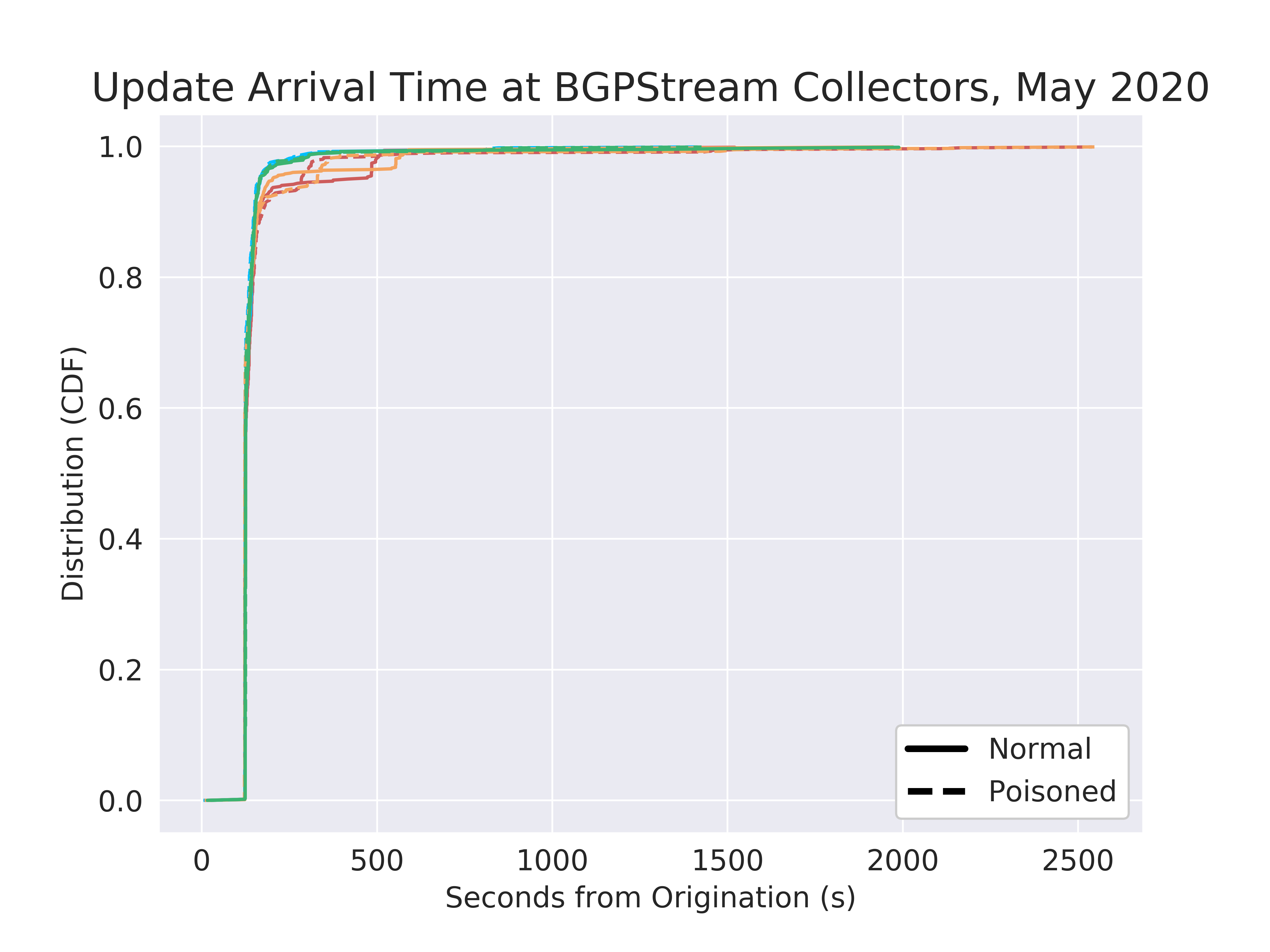}}
    \hspace{0.05cm}
    \caption{Update arrival time CDF. Each of five propagation experiments is illustrated in a different color.}
    \label{fig:propo}
\end{figure*}

\begin{figure*}[h]
    \centering
    \subfloat[August results.]{\label{fig:unique_aug}\includegraphics[width=0.45\textwidth]{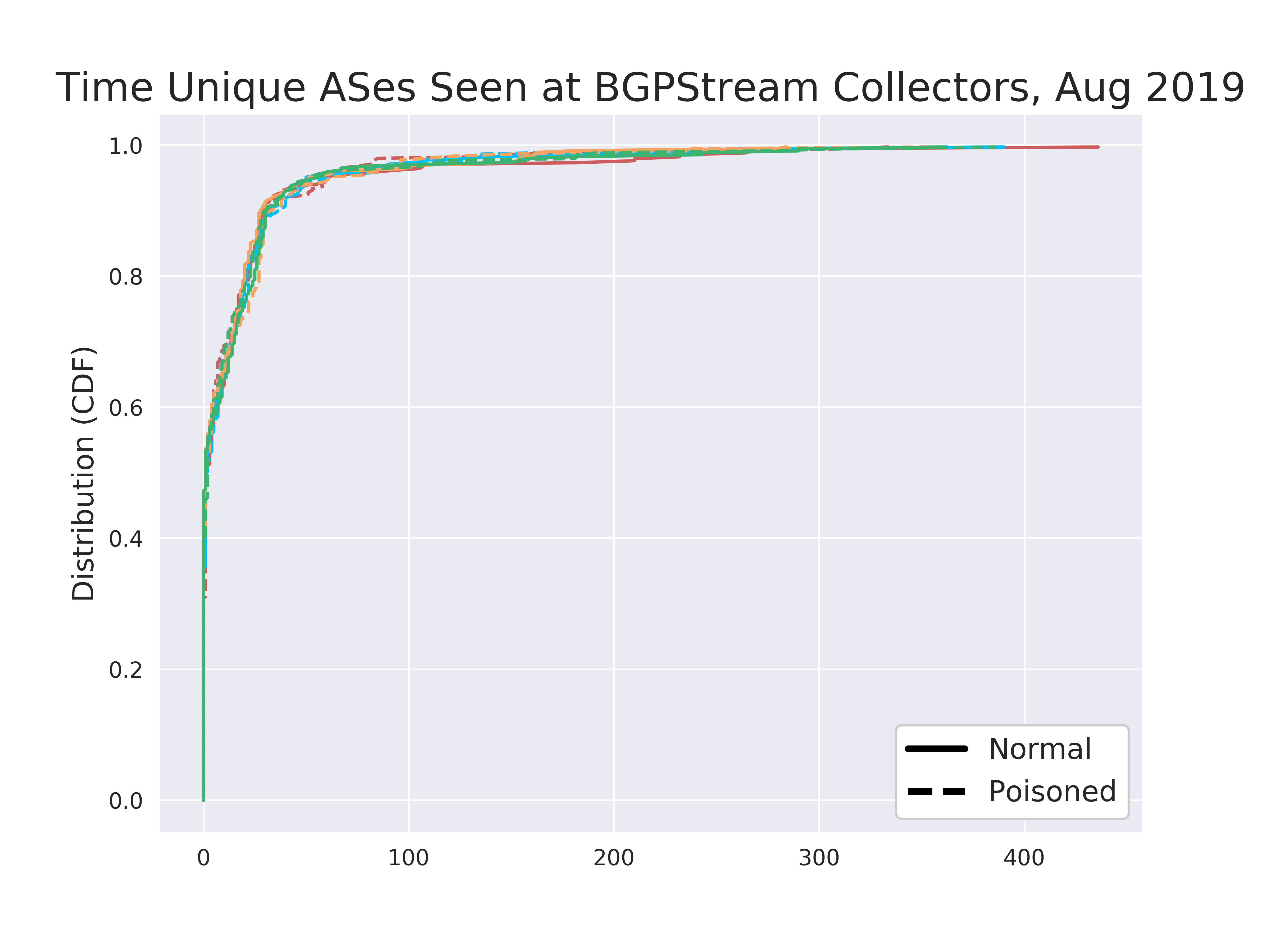}}
    \hspace{0.05cm}
    \subfloat[May results.]{\label{fig:unique_may}\includegraphics[width=0.45\textwidth]{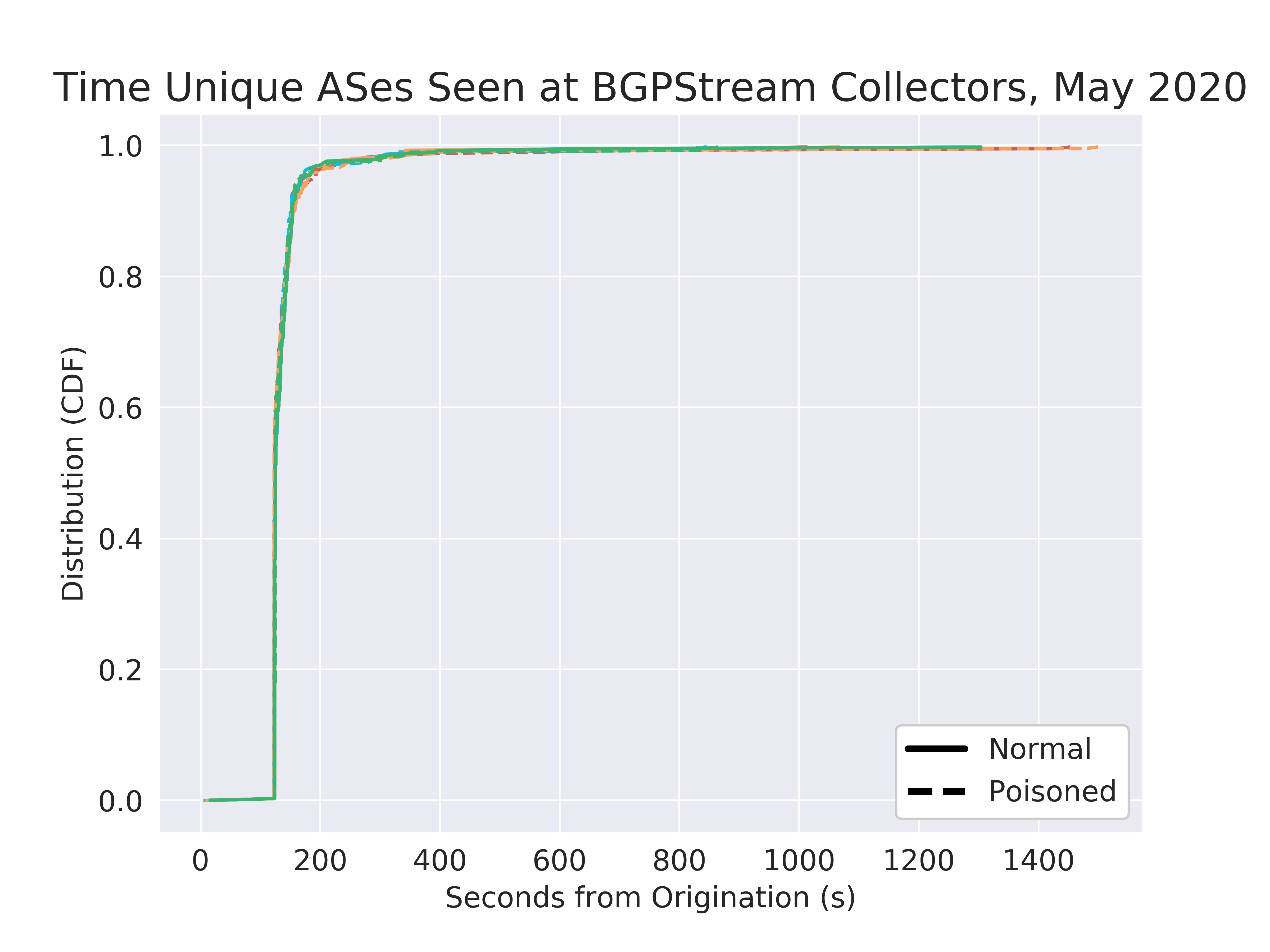}}
    \hspace{0.05cm}
    \caption{Unique AS arrival time CDF. Each of five propagation experiments is illustrated in a different color.}
    \label{fig:unique}
\end{figure*}

\end{document}